\documentclass[12pt]{article}                              
\usepackage{amsmath,amsthm,amsfonts,amscd,eucal,latexsym,amssymb}
\newlength{\dinwidth}
\newlength{\dinmargin}
\newsymbol\rest 1316         

\numberwithin{equation}{section}

\def\cA{{\cal A}}
\def\cB{{\cal B}}

\def\cD{{\cal D}}
\def\cE{{\cal E}}

\def\cH{{\cal H}}

\def\cN{{\cal N}}
\def\cO{{\cal O}}

\def\cP{{\cal P}}
\def\cQ{{\cal Q}}
\def\cR{{\cal R}}
\def\cS{{\cal S}}

\def\bC{{\mathbb C}}

\def\bN{{\mathbb N}}
\def\NN{{\mathbb N}}
\def\bR{{\mathbb R}}
\def\RR{{\mathbb R}}

\def\a{\alpha}
\def\b{\beta}
\def\g{\gamma}        \def\G{\Gamma}
\def\d{\delta}

\def\k{\kappa}
\def\l{\lambda}       \def\L{\Lambda}
\def\m{\mu}
\def\n{\nu}

\def\p{\pi}
\def\r{\rho}
\def\s{\sigma}

\def\t{\tau}

\def\o{\omega}        \def\O{\Omega}

\def\fV{{\mathfrak V}}

\def\fX{{\mathfrak X}}

\def\hh{\mathfrak h}

\newtheorem{Thm}{Theorem}[section]
\newtheorem{D+P}[Thm]{Definition and Proposition} 

\newtheorem{Prop}[Thm]{Proposition}
\newtheorem{Lemma}[Thm]{Lemma}
\theoremstyle{definition}
\newtheorem{Dfn}[Thm]{Definition}

\newtheorem{rem}[Thm]{Remark} 
\theoremstyle{remark}

\begin{document}
\newsymbol\bt 1202  
\newcommand{\lcrc}{\mbox{\footnotesize $\circ$}}
\newcommand{\Tmn}{T_{\mu\nu}}
\newcommand{\tmn}{\tau_{\mu\nu}}
\newcommand{\Ain}{{\cal A}_{\infty}}
\newcommand{\Cin}{C^{\infty}}
\newcommand{\Coin}{C^{\infty}_{0}}
\newcommand{\lb}{\mbox{\boldmath $[$}}
\newcommand{\rb}{\mbox{\boldmath $]$}}
\newcommand{\stern}{\star}
\newcommand{\at}{\alpha_{t}}
\newcommand{\kb}{\boldsymbol{k}}
\newcommand{\tb}{\boldsymbol{t}}
\newcommand{\kbs}{\boldsymbol{k}}
\newcommand{\tbs}{\boldsymbol{t}}
\newcommand{\xb}{\boldsymbol{\xi}}
\newcommand{\xbs}{\boldsymbol{\xi}}
\newcommand{\fv}{{\sf f}}
\newcommand{\gv}{{\sf g}}
\newcommand{\ku}{{\underline{k}}}
\newcommand{\xu}{{\underline{x}}}
\newcommand{\yu}{{\underline{y}}}
\newcommand{\xiu}{{\underline{\xi}}}
\newcommand{\Bu}{{\underline{B}}}
\newcommand{\bef}{\rhd}
\newcommand{\norm}[1]{\left\lVert #1 \right\rVert}
\newcommand{\betr}[1]{\left\lvert #1 \right\rvert}
\newcommand{\hs}[1]{\marginpar{$\Downarrow\Downarrow\Downarrow$}#1\marginpar{$\Uparrow\Uparrow\Uparrow$}}
\newcommand{\hide}[1]{} 
\newcommand{\ind}{\lhd}
\newcommand{\dni}{\rhd}
\newcommand{\dach}{\!\widehat{\;\;\;}}
\newcommand{\omi}{\text{\footnotesize $\Omega$}} 
\noindent
\begin{center}
{ \Large \bf
        Passivity and Microlocal Spectrum Condition}
\\[30pt]
{\large \sc Hanno Sahlmann {\rm and}  Rainer Verch}
\\[20pt]
                 Institut f\"ur Theoretische Physik,\\[4pt]
                 Universit\"at G\"ottingen,\\[4pt]
                 Bunsenstr.\ 9,\\[4pt]
                 D-37073 G\"ottingen, Germany\\[4pt]
                 e-mail: sahlmann$@$theorie.physik.uni-goettingen.de,\\
 verch$@$theorie.physik.uni-goettingen.de
\end{center}
${}$\\[26pt]
{\small {\bf Abstract. }
 In the setting of vector-valued quantum fields obeying a linear
 wave-equation in a globally hyperbolic, stationary spacetime, it is
 shown that the two-point functions of passive quantum states (mixtures of
 ground- or KMS-states) fulfill the microlocal spectrum condition
 (which in the case of the canonically quantized scalar field
 is equivalent to saying that the
 two-point function is of Hadamard form). The fields can be of bosonic
 or fermionic character. We also give an abstract
 version of this result by showing that passive states of a
 topological $*$-dynamical system have an asymptotic pair correlation
 spectrum of a specific type.
 }
${}$\\[10pt]
\section{Introduction}
\setcounter{equation}{0}
A recurrent theme in quantum field theory in curved spacetime is the
selection of suitable states which may be viewed as generalizations of
the vacuum state familiar from quantum field theory in flat
spacetime. The selection criterion for such states should, in
particular, reflect the idea of dynamical stability under temporal
evolution of the system. If a spacetime possesses a time-symmetry
group (generated by a timelike Killing vector field), then a ground
state with respect to the corresponding time-evolution appears as a
good candidate for a vacuum-like state. More generally, any thermal
equilibrium state for that time-evolution should certainly also be
viewed as a dynamically stable state. Ground- and thermal equilibrium
states, and mixtures thereof, fall into the class of the so-called
``passive'' states, defined in \cite{PW}. An important result by Pusz
and Woronowicz \cite{PW} asserts that a dynamical system is in a
passive state exactly if it is impossible to extract energy from the
system by means of cyclic processes. Since the latter form of
passivity, i.e.\ the validity of the second law of thermodynamics,
expresses a thermodynamical stability which is to be expected to hold
generally for physical dynamical systems, one would expect that
passive states are natural candidates for physical (dynamically
stable) states in quantum field theory in curved spacetime, at least
when the spacetime, or parts of it, posses time-symmetry groups. This
point of view has been expressed in \cite{BoBu}.

In this work we study the relationship between passivity of a quantum
field state and the microlocal spectrum condition for free quantum
fields on a stationary, globally hyperbolic spacetime.
The microlocal spectrum condition (abbreviated, $\m$SC\,) is a
condition restricting the form of the wavefront sets, WF$(\o_n)$, of
the $n$-point distributions $\o_n$ of a quantum field state
\cite{BFK,Rad1}. For quasifree states, it suffices to restrict the
form of WF$(\o_2)$; see relation \eqref{equ1.1} near the end of this
Introduction for a definition of $\m$SC in this case. There are
several reasons why the $\m$SC may rightfully be viewed as an
appropriate generalization of the spectrum condition (i.e.\ positivity
of the energy in any Lorentz frame), required for quantum fields in
flat spacetime, to quantum field theory in curved spacetime. Among the
most important is the proof by Radzikowski \cite{Rad1} (based on
mathematical work by Duistermaat and H\"ormander \cite{DH}) that, for
the free scalar Klein-Gordon field on any globally hyperbolic
spacetime, demanding that the two-point function $\o_2$ obeys the
$\m$SC is equivalent to $\o_2$ being of Hadamard form. This is
significant since it appears nowadays well-established to take the
condition that $\o_2$ be of Hadamard form as criterion for
phy\-sical (dynamically stable) quasifree states for linear quantum fields on
curved spacetime in view of a multitude of results, cf.\ e.g.\
\cite{Ful,FH,KW,Ver1,Ver.sym,Wald.Had,WaldII} and references given
therein.
Moreover, $\m$SC has several interesting structural properties which
are quite similar to those of the usual spectrum condition, and allow
to some extent
similar conclusions \cite{BFK,BF,V.acs}. It is particularly worth
mentioning that one may, in quasifree states of linear quantum fields
fulfilling $\m$SC, covariantly define Wick-products and develop the
perturbation theory for $P(\phi)_4$-type interactions along an
Epstein-Glaser approach generalized to curved spacetime
\cite{BF,BFK}. Also worth mentioning is the fact that $\m$SC
has proved useful in the analysis of other types of problems in
quantum field theory in curved spacetime \cite{Rad2,KRW,Few}.

In view of what we said initially about the significance of the
concept of passivity for quantum field states on stationary
spacetimes one would be inclined to expect that, on a stationary,
globally hyperbolic spacetime, a passive state fulfills the $\m$SC, at
least for quasifree states of linear fields. And this is what we are
going to establish in the present work.

We should like to point out that more special variants of such a
statement have been established earlier. For the scalar field obeying
the Klein-Gordon equation on a globally hyperbolic, static spacetime,
Fulling, Narcowich and Wald \cite{FNW} proved that the quasifree
ground state with respect to the static Killing vector field has a
two-point function of Hadamard form, and thus fulfills $\m$SC, as long
as the norm of the Killing vector field is globally bounded away from
zero. Junker \cite{Jun} has extended this result by showing that, if
the spacetime has additionally compact spatial sections, then the
quasifree KMS-states (thermal equilibrium states) at any finite
temperature fulfill $\m$SC. But the requirement of having compact
Cauchy-surfaces, or the constraint that the
static Killing vector field have a norm bounded globally away from
zero, exclude several interesting situations from applying the just
mentioned results. A prominent example is Schwarzschild spacetime,
which possesses a static timelike Killing flow, but the norm of the
Killing vector field tends to zero as one approaches the horizon along
any Cauchy-surface belonging to the static foliation. In \cite{Kay.HH}
(cf.\ also \cite{FulR}), quasifree ground- and KMS-states with respect
to the Killing flow on Schwarzschild spacetime have been constructed
for the scalar Klein-Gordon field, and it has long been conjectured
that the two-point functions of these states are of Hadamard
form. However, when trying to prove this along the patterns of
\cite{FNW} or \cite{Jun}, who use the formulation of quasifree
ground- and KMS-states in terms of the Klein-Gordon field's
Cauchy-data, one is faced with severe infra-red problems 
even for massive fields upon giving
up the constraint that the norm of the static Killing vector field be
globally bounded away from zero. This has called for trying to develop
a new approach to proving $\m$SC for passive states, the result of
which is our Theorem 5.1; see further below in this Introduction for a
brief description. As a corollary, our Thm.\ 5.1 shows that the
quasifree ground- and KMS-states of the scalar Klein-Gordon field on
Schwarzschild spacetime satisfy $\m$SC (thus their two-point functions
are of Hadamard form). [We caution the reader that this does not show
that these states or rather, their ``doublings'' defined in
\cite{Kay.HH}, were extendible to Hadamard states on the whole of the
Schwarzschild-Kruskal spacetime. There can be at most one single
quasifree, isometry-invariant Hadamard state on Schwarzschild-Kruskal
spacetime and this state necessarily restricts to a KMS-state at
Hawking temperature on the (``outer, right''--) Schwarzschild-part of
Schwarzschild-Kruskal spacetime, cf.\ \cite{KW,Kay.qf}.]

We turn to summarizing the contents of the present work. In Chapter 2,
we will introduce the notion of ``asymptotic pair correlation
spectrum'' of a state $\o$ of a topological $*$-dynamical system. This
object is to be viewed as a generalization of the wavefront set of the
two-point function $\o_2$ in the said general setting, see
\cite{V.acs} for further discussion. We then show that for (strictly)
passive states $\o$ the asymptotic pair correlation spectrum must be
of a certain, asymmetric form. This asymmetry can be interpreted as
the microlocal remnant of the asymmetric form of the spectrum that one
would obtain for a ground state.

Chapter 3 will be concerned with some aspects of wavefront sets of
distributions on test-sections of general vector bundles. Sec.\ 3.1
contains a reformulation of the wavefront set for vector-bundle
distributions along the lines of Prop.\ 2.2 in \cite{V.acs}. We
briefly recapitulate some notions of spacetime geometry, as far as
needed, in Sec.\ 3.2. In Sec.\ 3.3 we quote the propagation of
singularities theorem (PST) for wave-operators acting on vector
bundles, in the form used later in Chap.\ 5, from \cite{Den,DH}.

In Sec.\ 4.1 we introduce, following \cite{Keyl}, the Borchers algebra
of smooth test-sections with compact support in a vector bundle 
over a Lorentzian
spacetime, and briefly summarize the connection between states on the
Borchers algebra, their GNS-representations, the induced quantum
fields, and the Wightman $n$-point functions. We require that the
quantum fields associated with the states are, in a weak sense,
bosonic or fermionic, i.e.\ they fulfill a weak form of
(twisted) locality. A quite general formulation of (bosonic or
fermionic) quasifree states will be given in Sec.\ 4.3.

Chapter 5 contains our main result, saying that for a state $\o$ on
the Borchers algebra associated with a given vector bundle, over a globally
hyperbolic, stationary spacetime $(M,g)$ as base manifold,
the properties
\begin{itemize}
\item[(i)] $\o$ is (strictly) passive,
\item[(ii)] $\o$ fulfills a weak form of (twisted) locality, and
\item[(iii)] $\o_2$ is a bi-solution up to $C^{\infty}$ for a wave
  operator,
\end{itemize}
imply
\begin{equation}
\label{equ1.1}
 {\rm WF}(\o_2) \subset \cR\,,
\end{equation}
where $\cR$ is the set of pairs of non-zero covectors $(q,\xi;q',\xi') \in {\rm
  T}^*M \times {\rm T}^*M$ so that $g^{\m \n}\xi_{\n}$ is
past-directed and lightlike, the base points $q$ and $q'$ are
connected by an affinely parametrized, lightlike geodesic $\g$, 
and both $\xi$ and $-\xi'$ are co-tangent to $\g$,
or $\xi = - \xi'$ if $q = q'$.

Following \cite{BFK}, we say that the quasifree state with two-point
function $\o_2$ fulfills the $\m$SC if the inclusion \eqref{equ1.1}
holds. If one had imposed the additional requirement that $\o$ (resp.,
the associated quantum fields) fulfill appropriate vector-bundle
versions of the CCR or CAR, one would conclude that 
$$ {\rm WF}(\o_2) = \cR\,,$$
as is e.g.\ the case for the free scalar Klein-Gordon field (cf.\ \cite{Rad1}).
Moreover, for a quasifree state $\o$ on the Borchers algebra of a
vector bundle over any globally hyperbolic spacetime one can show that
imposing CCR or CAR implies that $\o_2$ is of Hadamard form
(appropriately generalized) if and only if WF$(\o_2) = \cR$. The
discussion of these matters will be contained in a separate article
\cite{SaV}.


\section{Passivity and Asymptotic Pair Correlation Spectrum}
\setcounter{equation}{0}
Let $\cA$ be a $C^*$-algebra with unit and $\{\at\}_{t \in \bR}$ a
one-parametric group of automorphisms of $\cA$, supposed to be
strongly continuous, that is, $||\,\at(A) - A\,|| \to 0$ as $t \to 0$
for each $A \in \cA$. Moreover, let $D(\d)$ denote the set of all $A
 \in \cA$ such that the limit
$$ \d(A) := \lim_{t \to 0}\,\frac{1}{t}(\at(A)- A) $$
exists. One can show that $D(\d)$ is a dense $*$-subalgebra of $\cA$,
and $\d$ is a derivation with domain $D(\d)$.

Following \cite{PW}, one calls a state $\omega$ on $\cA$ {\it
  passive} if for all unitary elements $U \in D(\d)$ which are
continuously connected to the unit element,\,\footnote{i.e.\ there
  exists a continuous curve $[0,1] \owns t \mapsto U(t) \in D(\d)$
  with each $U(t)$ unitary and $U(0) = 1_{\cA}$, $U(1) = U$.}
the estimate 
\begin{equation}
\label{equ4}
 \omega(U^*\frac{1}{i}\d(U)) \ge 0
\end{equation}
is fulfilled. As a consequence, $\omega$ is invariant under
$\{\at\}_{t\in\bR}$: $\omega \lcrc \at = \omega$ for all $t \in
\bR$. Furthermore, it can be shown (cf.\ \cite{PW}) that ground
states or KMS-states at inverse temperature $\b \ge 0$ for $\at$ are
passive, as are convex sums of such states. (In Appendix A we will
summarize some basic properties of ground states and KMS-states.
Standard references include \cite{BR2,Sak}.)

However, the significance of passive states is based on two remarkable
results in \cite{PW}. First, a converse of the previous statement is
proven there: If a state is completely passive, then it is a ground
state or a KMS-state at some inverse temperature $\b \ge 0$. Here a
state is called {\it completely passive} if, for each $n \in \bN $,
the product state $\otimes^n\omega$ is a passive state on
$\otimes^n\cA$ with respect to the dynamics
$\{\otimes^n\at\}_{t\in\bR}$.

Secondly, the following is established in \cite{PW}: the dynamical
system modelled by $\cA$ and $\{\at\}_{t \in\bR}$ is in a passive state
precisely if it is impossible to extract energy from the system by
means of cyclic processes. In that sense, passive states may be viewed
as good candidates for physically realistic states of any dynamical
system since for these states the second law of thermodynamics is
warranted.

 In the present section we
are interested in studying the asymptotic high frequency behaviour of
passive states along similar lines as developed recently in
\cite{V.acs}.
We shall, however, generalize the setup since this will prove useful
for developments later in this work. Thus, we assume now that
$\cA$ is a topological $*$-algebra with a locally convex topology and
with a unit element (cf.\ e.g.\ \cite{Sch}). We
denote by $S$ the set of continuous semi-norms for $\cA$.

Moreover, we say that $\{\at\}_{t\in\bR}$ is a continuous
one-parametric group of $*$-auto\-mor\-phisms of $\cA$ if for each $t$,
$\at$ is a topological $*$-automorphism of $\cA$, and if the group
action is locally bounded and
continuous in the sense that for each $\s \in S$ there is $\s' \in S$,
$r > 0$ with $\s(\at(A)) \le \s'(A)$ for all $|t| < r$, $A \in \cA$, and
$\s(\at(A) - A) \to 0$ as $t \to 0$ for
each $A \in \cA$. Then we refer to the pair $(\cA,\{\at\}_{t\in\bR})$
as a {\it topological $*$-dynamical system}. Using the fact that for
all $A,B \in \cA$ and $\s \in S$, the maps $C \mapsto \s(AC)$ and $C
\mapsto \s(CB)$ are again continuous semi-norms on $\cA$, one deduces
by a standard argument that also $\s(\a_s(A)\a_t(B) - AB) \to 0$ as
$s,t \to 0$.

 A continuous linear functional $\omega$ on $\cA$
will be called a state if $\omega(A^*A) \ge 0$ for all $A \in \cA$ and if
$\omega(1_{\cA}) = 1$. Furthermore, we say that $\omega$ is a ground
state, or a KMS-state at inverse temperature $\b > 0$, for 
$\{\at\}_{t\in\bR}$, if the functions $t \mapsto \omega(A\at (B))$ are
bounded for all $A,B \in \cA$, and if $\omega$ satisfies the ground
state condition \eqref{equ2} or the KMS-condition \eqref{equ3} given
in Appendix A, respectively.

Now we call a family $(A_{\l})_{\l > 0}$ with $A_{\l} \in \cA$ a {\it
  global testing family}  in
$\cA$ provided there is for each $\s\in S$ an $s \ge 0$ (depending on
$\s$ and on the family) such that 
\begin{equation}
\label{equ6}
  \sup_{\l}\, \l^s\s(A_{\l}^*A_{\l}) < \infty\,.
\end{equation}
The set of all global testing families will be denoted by ${\bf A}$.

Let $\omega$ be a state on $\cA$, and $\xb = (\xi_1,\xi_2) \in
\bR^2\backslash\{0\}$. Then we say that $\xb$ is a {\it regular
  direction} for $\omega$, with respect to the continuous
one-parametric group $\{\at\}_{t \in \bR}$, if there exists some $h
\in \Coin(\bR^2)$ and an open neighbourhood $V$ of $\xb$ in $\bR^2
\backslash \{0\}$ such that\,\footnote{We
  shall write $\varphi(\l) = O^{\infty}(\l)$ as $\l \to 0$ iff for
  each $s \in \bN$ there are $C_s, \l_s > 0$ so that $|\varphi(\l)|
  \le C_s\,\l^s$ for all $0 < \l < \l_s$.}
\begin{equation}
\sup_{\kb \in V}\, \left| \int {\rm e}^{-i\l^{-1}\kb\cdot
    \tbs}h(\tb) \omega(\a_{t_1}(A_{\l})\a_{t_2}(B_{\l}))\,d\tb \right|
= O^{\infty}(\l) \quad {\rm as}\ \l \to 0
\end{equation}
holds for all global testing families $(A_{\l})_{\l > 0},(B_{\l})_{\l
  > 0} \in {\bf A}$. 

Then we define the set ${\it ACS}^2_{\bf A}(\omega)$ as the complement
in $\bR^2\backslash\{0\}$ of all
 $\kb$ which are regular directions for $\omega$.
We call ${\it ACS}{}^2_{\bf A}(\omega)$ the {\it global asymptotic pair
  correlation spectrum} of $\omega$. The asymptotic pair correlation
spectrum, and more generally, asymptotic $n$-point correlation spectra
of a state, may be regarded as generalizations of the notion of
wavefront set of a distribution in the setting of states on a
dynamical system. We refer to \cite{V.acs} for considerable further
discussion and motivation. The properties of ${\it
  ACS}{}^2_{\bf A}(\omega)$ are analogous to those of ${\it
  ACS}^2(\omega)$ described in \cite[Prop.\ 3.2]{V.acs}. In
particular, ${\it ACS}{}^2_{\bf A}(\omega)$ is a closed conic set in
$\bR^2\backslash \{0\}$.
 It is evident that, if $\o$ is a finite convex sum of states $\o_i$, then
 ${\it ACS}^2_{\bf A}(\o)$ is contained in $\bigcup_i{\it ACS}^2_{\bf
   A}(\o_i)$.
  
Now we are going to establish an upper bound for ${\it
  ACS}{}^2_{\bf A}(\omega)$, distinguished by a certain asymmetry, for
all $\omega$ in a subset $\cP$ of the set of all passive states, to be defined
next:

We define $\cP$ as the set of all states on $\cA$ which are of the
form
\begin{equation}
\omega(A) = \sum_{i = 1}^m \r_i \omega_i(A)\,, \quad A \in \cA\,,
\end{equation}
where $m \in \bN$, $\r_i > 0$, $\sum_{i =1}^m \r_i = 1$, and each
$\omega_i$ is a ground state or a KMS-state at some inverse temperature
$\b_i > 0$ (note that $\b_i = 0$ is {\it not} admitted!) on $\cA$ with
respect to $\{\at\}_{t \in \bR}$. The states in $\cP$ will be called
{\it strictly passive}.

We should like to remark that in the present general setting where
$\cA$ is not necessarily a $C^*$-algebra, the criterion for passivity 
given at the beginning in \eqref{equ4} may be inappropriate since it could
happen that $D(\d)$, even if dense in $\cA$, doesn't contain
sufficiently many unitary elements. In the $C^*$-algebraic situation,
\eqref{equ4} entails the slightly weaker variant
\begin{equation}
\label{equ5}
 \omega(A\frac{1}{i}\d(A)) \ge 0
\end{equation}
for all $A = A^* \in D(\d)$, and one may take this as substitute for
the condition of passivity of a state in the present more general
framework (supposing that $D(\d)$ is dense). In fact, each $\omega \in
\cP$ is $\{\at\}_{t\in\bR}$-invariant and
 satisfies \eqref{equ5} (see Appendix \ref{app1}), and in the $C^*$-algebraic
situation, every $\omega \in \cP$ also satisfies \eqref{equ4}. 
\begin{Prop} 
\label{pro1}
Let $(\cA,\{\at\}_{t\in\bR})$ be a topological
  $*$-dynamical system as described above. 
\begin{itemize}
\item[{\rm (1)}] Let $\omega \in \cP$. Then\\[4pt]
       either \quad ${\it ACS}^2_{\bf A}(\omega) = \emptyset$, \\[4pt]
       or \quad \ ${\it ACS}^2_{\bf A}(\omega) = \{(\xi_1,\xi_2) \in
       \bR^2\backslash \{0\} : \xi_1 + \xi_2 = 0,\ \xi_2 \ge 0\}$.
\item[{\rm (2)}] Let $\omega$ be an $\{\at\}_{t\in\bR}$-invariant
  KMS-state at inverse temperature $\b = 0$. Then\\[4pt]
       either \quad ${\it ACS}^2_{\bf A}(\omega) = \emptyset$, \\[4pt]
       or \quad \ ${\it ACS}^2_{\bf A}(\omega) = \{(\xi_1,\xi_2) \in
       \bR^2\backslash \{0\} : \xi_1 + \xi_2 = 0\}$.
\end{itemize}
\end{Prop}
\begin{proof}
{\it 1.) }  By assumption $\omega$ is continuous, hence we can find a
seminorm $\s \in S$ so that $|\omega(A)| \le \s(A)$ for all $A \in
\cA$. Thus there are positive constants $c$ and $s$ so that
\begin{equation}
\label{equ13}
\omega(\at(A^*_{\l}A_{\l})) = \omega(A^*_{\l}A_{\l}) \le c
\cdot(1+\l^{-1})^{s}
\end{equation}
holds for all $t\in\bR$. In the first equality, the invariance of
$\omega$ was used, and in the second, condition \eqref{equ6} was applied.
Thus, for any Schwartz-function $\hat{h} \in \cS(\bR^2)$,
 and any $(A_{\l})_{\l >
  0}$, $(B_{\l})_{\l > 0}$ in ${\bf A}$, one obtains that the following
function of $\l > 0$ and $\kb \in \bR^2$,
$$ w_{\l}(\kb) := \int {\rm e}^{-i\kb \cdot
  \tb}\hat{h}(\tb)\omega(\a_{t_1}(A_{\l})\a_{t_2}(B_{\l}))\,d\tb
$$
depends smoothly on $\kb$ and satisfies the estimate
$$ |w_{\l}(\l^{-1}\kb)| \le c'(|\kb| +\l^{-1} + 1)^r $$
with suitable constants $c' > 0$, $r \in \bR$. Hence, this function
satisfies the assumptions of Lemma 2.2 in \cite{V.acs}. Application of
the said Lemma entails the following: Suppose that for some open
neighbourhood $V$ of $\xb \in \bR^2\backslash\{0\}$ we can find some
$\hat{h} \in \cS(\bR^2)$ with $\hat{h}(0) = 1$ and 
\begin{equation}
\label{equ7}
 \sup_{\kb \in V}\,   \left|   \int {\rm e}^{-i\l^{-1}\kb \cdot
  \tb}\hat{h}(\tb)\omega(\a_{t_1}(A_{\l})\a_{t_2}(B_{\l}))\,d\tb \right| =
O^{\infty}(\l) \quad {\rm as} \ \l \to 0
\end{equation}
for all $(A_{\l})_{\l>0},(B_{\l})_{\l>0} \in {\bf A}$. Then this
implies that the analogous relation holds with $\hat{h}$ replaced by $\phi
\cdot \hat{h}$ for any $\phi \in \Coin(\bR^2)$ when simultaneously $V$ is
replaced by some slightly smaller neighbourhood $V'$ of
$\xb$. Consequently, relation \eqref{equ7} --- with $\hat{h} \in
\cS(\bR^2)$, $\hat{h}(0)= 1$ --- entails that $\xb$ is absent 
from ${\it ACS}^2_{\bf  A}(\omega)$.
\\[6pt]
{\it 2.) } Some notation needs to be introduced before we can proceed.
For $f \in \cS(\bR)$, we define
$$ (\t_s f)(s') := f(s' -s) \quad {\rm and} \quad {}^r\!\!f(s') :=
f(-s')\,, \quad s,s' \in \bR\,.
$$
Then we will next establish
\begin{equation}
\label{equ8}
\omega \lcrc \at = \omega \quad \Rightarrow \quad {\it ACS}^2_{\bf A}(\omega)
\subset \{(\xi_1,\xi_2) \in \bR^2\backslash\{0\}: \xi_1 + \xi_2 = 0\}\,.
\end{equation}
To this end, let $\xb = (\xi_1,\xi_2) \in \bR^2\backslash\{0\}$ be
such that $\xi_1 + \xi_2 \ne 0$, and pick some $\d > 0$ and an open
neighbourhood $V_{\xbs}$ of $\xb$ so that $|k_1 + k_2| > \d$ for all
$\kb \in V_{\xbs}$.

Now pick two functions ${h}_j \in \Coin(\bR)$ $(j =1,2)$ such
that their Fourier-transforms $\hat{h}_j(t_j) = \frac{1}{\sqrt{2\p}}\int
{\rm e}^{-it_j\cdot p}{h}_j(p)\,dp$ have the property $\hat{h}_j(0) =
1$. Define $\hat{h} \in \cS(\bR^2)$ by $\hat{h}(\tb) := \hat{h}_1(t_1)
\hat{h}_2(t_2)$.

Then observe that one can find $\l_0 > 0$ such that the functions
\begin{equation}
\label{equ10}
{g}_{\l,\kbs}(p) := (\,(\t_{-\l^{-1}(k_1 +
  k_2)}{}^r\!{h}_1)\cdot {h}_2\,)(p)\,, \quad p \in
\bR\,,
\end{equation}
vanish for all $\kb = (k_1,k_2) \in V_{\xbs}$ and all $0 < \l < \l_0$.
Consequently, also the functions
\begin{equation}
\label{equ9}
 {f}_{\l,\kbs}(p) := (\t_{\l^{-1}k_2}{g}_{\l,\kbs})(p)\,,
 \quad p \in \bR\,,
\end{equation}
vanish for all $\kb \in V_{\xbs}$ and all $0 < \l < \l_0$. Denoting
the Fourier-transform of ${f}_{\l,\kbs}$ by $\hat{f}_{\l,\kbs}$, one
obtains for all $\kb \in V_{\xbs}$, $0 < \l < \l_0$:
\begin{eqnarray*}
 0 &=&\int \hat{f}_{\l,\kbs}(s)\,\omega(A_{\l}\a_s(B_{\l}))\,ds \\
  & = & \int {\rm e}^{-i\l^{-1}(k_1 + k_2)s'}{\rm
    e}^{-i{\l}^{-1}k_2s}\hat{h}_1(s')\hat{h}_2(s' +
  s)\omega(A_{\l}\a_s(B_{\l}))\,ds'\,ds\\
 & = & \int {\rm
   e}^{-i\l^{-1}\kbs\cdot\tbs}\hat{h}(\tb)\,\omega(\a_{t_1}(A_{\l})\a_{t_2}
 (B_{\l}))\,
d\tb 
\end{eqnarray*}
for all testing-families $(A_{\l})_{\l > 0}$, $(B_{\l})_{\l >
  0}$. Invariance of $\omega$ under $\{\at\}_{t\in\bR}$ was used in
passing from the second equality to the last. 
In view of step {\it 1.)} above, this shows \eqref{equ8}.
\\[6pt]
{\it 3.) } In a further step we will argue that 
\begin{equation}
\label{equ11}
 \omega\ {\rm ground \ state}  \quad \Rightarrow \quad 
 {\it ACS}^2_{\bf A}(\omega) \subset \{(\xi_1,\xi_2) \in
 \bR^2\backslash\{0\}: \xi_2 \ge 0\}\,.
\end{equation}
So let again ${h}_j$ and $\hat{h}_j$ as above, and
${f}_{\l,\kbs}$ as in \eqref{equ9} with Fourier-transform
$\hat{f}_{\l,\kbs}$. Let $\xb = (\xi_1,\xi_2) \in \bR^2\backslash\{0\}$ have
$\xi_2 < 0$. Then there is an open neighbourhood $V_{\xbs}$ of $\xb$
and an $\epsilon > 0$ so that $k_2 < -\epsilon$ for all $\kb \in
V_{\xbs}$. The support of ${f}_{\l,\kbs}$ is contained in the
support of $\t_{\l^{-1}k_2}{h}_2$, and there is clearly some
$\l_0 > 0$ such that ${\rm supp}\,\t_{\l^{-1}k_2}{h}_2 \subset
(-\infty,0)$ for all $\kb = (k_1,k_2) \in V_{\xbs}$ as soon as $0 < \l
< \l_0$.  By the
characterization of a ground state given in \eqref{equ2}, and using also the
$\{\at\}_{t\in\bR}$-invariance of a ground state, one therefore
obtains
\begin{eqnarray*}
\lefteqn{ \sup_{\kbs \in V_{\xbs}}\,
\left|   \int {\rm e}^{-i\l^{-1}\kbs \cdot
  \tbs}\hat{h}(\tb)\omega(\a_{t_1}(A_{\l})\a_{t_2}(B_{\l}))\,d\tb \right|
}\\
 & = & \sup_{\kbs \in V_{\xbs}}\, \left| \int
\hat{f}_{\l,\kbs}(s)\omega(A_{\l}\a_s(B_{\l}))\,ds \right|
\\
& = & 0 \quad {\rm if} \ \, 0 < \l < \l_0
\end{eqnarray*}
for all $(A_{\l})_{\l > 0},(B_{\l})_{\l > 0} \in {\bf A}$. Relation
\eqref{equ11} is thereby proved.
\\[6pt]
{\it 4.) } Now we turn to the case 
\begin{equation}
\label{equ14}
\omega \ {\rm KMS\ at}\ \b > 0 \quad \Rightarrow \quad
 {\it ACS}^2_{\bf A}(\omega)
\subset \{(\xi_1,\xi_2) \in \bR^2\backslash\{0\} : \xi_2 \ge 0\}\,.
\end{equation}
Consider a $\xb \in \bR^2\backslash\{0\}$ with $\xi_2 < 0$ and pick
some $\epsilon > 0$ and an open neighbourhood $V_{\xb}$ of $\xb$ so
that $k_2 < -\epsilon$ for all $\kb = (k_1,k_2) \in V_{\xb}$. Choose
again ${h}_j$ and $\hat{h}_j$ as above and define correspondingly
${g}_{\l,\kb}$ and ${f}_{\l,\kb}$ as in \eqref{equ10} and
\eqref{equ9},
respectively. Denote again their Fourier-transforms by $\hat{g}_{\l,\kb}$ and
$\hat{f}_{\l,\kb}$. Note that ${g}_{\l,\kb}$ and ${f}_{\l,\kb}$
are in $\Coin(\bR)$ for all $\l > 0$ and all $\kb \in \bR^2$, so their
Fourier-transforms are entire analytic. Moreover, a standard estimate
shows that 
\begin{equation}
\label{equ12}
\sup_{\l>0,\kb \in \bR^2}\,\int |\hat{g}_{\l,\kb}(s + i\b)|\,ds \le c' <
\infty\,.
\end{equation}
One calculates
$$ \hat{f}_{\l,\kb}(s + i\b) = {\rm e}^{\l^{-1}k_2\b}{\rm
  e}^{-i\l^{-1}k_2s}\hat{g}_{\l,\kb}(s + i\b)\,, \quad s \in \bR\,,$$
and now the KMS-condition \eqref{equ3} yields for all $(A_{\l})_{\l >
  0},(B_{\l})_{\l > 0} \in {\bf A}$,
\begin{eqnarray*}
\lefteqn{\left | \int \hat{f}_{\l,\kb}(s)\omega(A_{\l}\a_s(B_{\l}))\,ds\right|} \\
& = & \left| {\rm e}^{\l^{-1}k_2\b}\int{\rm
    e}^{-i\l^{-1}k_2s}\hat{g}_{\l,\kb}(s +
  i\b)\omega(\a_s(B_{\l})A_{\l})\,ds \right| \\
& \le & {\rm e}^{\l^{-1}k_2\b}c'\cdot c''(1+\l^{-1})^{s'}\,, \quad \l > 0,\
\kb \in \bR^2\,,
\end{eqnarray*}
for suitable $c'',s' > 0$, where \eqref{equ13} and \eqref{equ12} have been
used. Making also use of the $\{\at\}_{t\in\bR}$-invariance of
$\omega$ one finds, with suitable $\g > 0$,
\begin{eqnarray*}
\lefteqn{ \sup_{\kb \in V_{\xb}}\,
\left|   \int {\rm e}^{-i\l^{-1}\kbs \cdot
  \tbs}\hat{h}(\tb)\omega(\a_{t_1}(A_{\l})\a_{t_2}(B_{\l}))\,d\tb \right|
}\\
 & = & \sup_{\kbs \in V_{\xbs}}\, \left| \int
\hat{f}_{\l,\kbs}(s)\omega(A_{\l}\a_s(B_{\l}))\,ds \right|
\\
& \le & \g\,{\rm e}^{-\l^{-1}\epsilon\b}(1+\l^{-1})^{s'} = O^{\infty}(\l)
\quad {\rm as}\ \l \to 0
\end{eqnarray*}
for all $(A_{\l})_{\l >
  0},(B_{\l})_{\l > 0} \in {\bf A}$. This establishes statement \eqref{equ14}.
\\[6pt]
{\it 5.) } Combining now the assertions \eqref{equ8}, \eqref{equ11} and
\eqref{equ14}, one can see
that for each $\omega \in \cP$ there holds
$$ {\it ACS}^2_{\bf A}(\omega) \subset \{(\xi_1,\xi_2) \in
\bR^2\backslash\{0\}: \xi_1 + \xi_2 = 0,\ \xi_2 \ge 0\}\,.$$
Since the set on the right-hand side obviously has no proper conic
subset in $\bR^2\backslash\{0\}$, one concludes that statement (1) of
the Proposition holds true.
\\[6pt]
{\it 6.) }
As $\omega$ is KMS at $\b = 0$, this means that it is a trace:
$\omega(AB) = \omega(BA)$. Since $\omega$ is also
$\{\at\}_{t\in\bR}$-invariant, we have 
$$ {\it ACS}^2_{\bf A}(\omega) \subset \{(\xi_1,\xi_2) \in
\bR^2\backslash\{0\}: \xi_1 + \xi_2 =0\}\,.$$
The set on the right hand side has precisely two proper closed conic
subsets
$$ W_{\pm}:= \{(\xi_1,\xi_2) \in \bR^2\backslash\{0\}: \xi_1 +\xi_2 = 0,\
\pm \xi_2 \ge 0\}\,.
$$
These two sets are disjoint, $W_{+} \cap W_{-} = \emptyset$,
and we have $W_{+} = -W_{-}$. Hence, since $\omega$ is a
trace, one can argue exactly as in \cite[Prop.\ 4.2]{V.acs} to
conclude that either ${\it ACS}^2_{\bf A}(\omega) \subset W_{+}$ or
${\it ACS}^2_{\bf A}(\omega) \subset W_{-}$ imply ${\it ACS}^2_{\bf
  A}(\omega) = \emptyset$. This establishes statement (2) of the Proposition.
\end{proof}
Hence we see that strict passivity of $\omega$ results in its ${\it
  ACS}^2_{\bf A}(\omega)$ being asymmetric. This is due to the fact
that, roughly speaking, the negative part of the spectrum of the
unitary group implementing $\{\at\}_{t \in \bR}$ in such a state is
suppressed by an exponential weight factor. It is worth noting that
this asymmetry is not present for KMS-states at $\b = 0$. Such states
at infinite temperature would hardly be regarded as candidates for
physical states, and they can be ruled out by the requirement that
${\it ACS}^2_{\bf A}(\omega)$ be asymmetric.  
\begin{rem}
\label{rem1}
 One can modify or, effectively, enlarge the set of testing
families by allowing a testing family to depend on additional
parameters: Define ${\bf A^{\sharp}}$ as the set of all families
$(A_{y,\l})_{\l > 0,y\in\bR^m}$ where $m \in \bN$ is arbitrary (and
depends on the family) having the property that for each semi-norm $\s
\in S$ there is an $s \ge 0$ (depending on $\s$ and on the family)
such that
\begin{equation}
 \sup_{\l,y}\, \l^s\s(A_{y,\l}^*A_{y,\l}) < \infty\,.
\end{equation}
Then the definition of a regular direction $\kb \in
\bR^2\backslash\{0\}$ for a state $\o$ of the dynamical system
$(\cA,\{\at\}_{t\in\bR})$ may be altered through declaring $\xb$ a
regular direction iff there are an open neighbourhood $V$ of $\xb$
and a function $h \in \Coin(\bR^2)$, $h(0) = 1$, so that
$$ \sup_{\kb \in V}\,\sup_{y,z}\,
\left| \int {\rm e}^{-i\l^{-1}\kb\cdot
    \tb}h(\tb)\o(\a_{t_1}(A_{y,\l})\a_{t_2}(B_{z,\l}))\, d\tb \right| =
O^{\infty}(\l) \quad {\rm as}\ \l \to 0
$$
holds for any pair of elements $(A_{y,\l})_{\l>0,y\in\bR^m}$,
$(B_{z,\l})_{\l>0,z\in\bR^n}$ in ${\bf A}^{\sharp}$. This makes the set of
regular directions a priori smaller, and if we define ${\it
  ACS}^2_{{\bf A}^{\sharp}}(\o)$ as the complement of all $\xb \in
\bR^2\backslash\{0\}$ that are regular directions for $\o$ according
to the just given, altered definition then clearly we have, in general,
${\it  ACS}^2_{{\bf A}^{\sharp}}(\o) \supset{\it
  ACS}^2_{{\bf A}}(\o)$. However, essentially by repeating --- with
somewhat more laborious notation --- the proof of Prop.\ \ref{pro1}, one can
see that the statements of Prop.\ \ref{pro1} remain valid upon replacing 
${\it  ACS}^2_{{\bf A}}(\o)$ by ${\it
  ACS}^2_{{\bf A}^{\sharp}}(\o)$. We shall make use of that observation
later.
\end{rem}

\section{Wavefront Sets and Propagation of Singularities}
\label{sec2}
\setcounter{equation}{0}
\subsection{Wavefront Sets of Vectorbundle-Distributions}

Let $\fX$ be a $\Cin$ vector bundle over a base manifold $N$ ($n$ =
dim\,$N \in \bN$) with typical fibre isomorphic to $\bC^{\,r}$ or
 to $\bR^{\,r}$; the bundle
projection will be denoted by $\p_N$. (We note that here and
throughout the text, we take manifolds to be $C^{\infty}$, Hausdorff,
2nd countable, finite dimensional and without boundary.)
 We shall write $\Cin(\fX)$ for
the space of smooth sections of $\fX$ and $\Coin(\fX)$ for the
subspace of smooth sections with compact support. These spaces can be
endowed with locally convex topologies in a like manner as for the
corresponding test-function spaces $\cE(\bR^n)$ and $\cD(\bR^n)$, cf.\
\cite{Dieu3,Dieu7} for details. By $(\Cin(\fX))'$
and $(\Coin(\fX))'$ we denote the respective spaces of continuous
linear functionals, and by $\Coin(\fX_U)$ the space of all smooth
sections in $\fX$ having compact support in the open subset $U$ of $N$.

For later use, we introduce the following terminology.
  We say that $\r$ is a local diffeomorphism of some manifold $X$ if $\r$ is
  defined on some open subset $U_1 = {\rm dom}\,\r$ of $X$ and maps it
  diffeomorphically onto another open subset $U_2 = {\rm Ran}\,\r$ of
  $X$. If $U_1 =  U_2 = X$, then $\r$ is a diffeomorphism as usual.
Let $\r$ be a (local) diffeomorphism of the base manifold $N$.
Then we say that $R$ is a (local) {\it bundle map of} $\fX$
{\it covering} $\r$ if $R$ is
a smooth map from $\p_N^{-1}({\rm dom}\,\r)$ to $\p_N^{-1}({\rm
  Ran}\,\r)$ so that, for each $q$ in ${\rm dom}\,\r$, $R$ maps the
fibre over $q$ linearly into the fibre over $\r(q)$. If this map is
also one-to-one and if $R$ is also a local diffeomorphism, then $R$
will be called a (local) {\it morphism of} $\fX$ {\it covering} $\r$.
Moreover, let $(\r_x)_{x \in B}$ be a family of (local)
diffeomorphisms of $N$ depending smoothly on $x \in B$ where $B$ is an
open neighbourhood of $0 \in \bR^s$ for some $s \in \bN$. Then we call
 $(R_x)_{x \in B}$  a {\it family of (local) morphisms of} $\fX$
{\it covering} $(\r_x)_{x\in B}$ if each $R_x$ is a morphism of $\fX$
covering $\r_x$, depending smoothly on $x \in B$.

Note that each bundle map $R$ of $\fX$ covering a (local)
differomorphism $\r$ of $N$ induces a (local) action on $\Coin(\fX)$
in form of a continuous linear map $R^{\stern} : \Coin(\fX_{{\rm
    dom}\,\r}) \to \Coin(\fX_{{\rm Ran}\,\r})$ given by
\begin{equation}
 R^{\stern}f := R \lcrc f \lcrc \r^{-1}\,, \quad f \in \Coin(\fX_{{\rm
     dom}\,\r})\,.
\end{equation}

Given a local trivialization of $\fX$ over some open $U \subset N$,
this induces a one-to-one correspondence between $\Coin(\fX_U)$ and
$\oplus^r\cD(U)$, inducing in turn a one-to-one correspondence between
$(\Coin(\fX_U))'$ and $\oplus^r\cD'(U)$. Now let $u
\in(\Coin(\fX_U))'$ and let $(u_1,\ldots,u_r) \in \oplus^r \cD'(U)$ be
the corresponding $r$-tupel of scalar distributions on $U$ induced by
the local trivialization of $\fX$ over $U$. The wavefront set WF$(u)$
of $u \in (\Coin(\fX_U))'$ may then be defined as the union of the
wavefront sets of the components $u_a$, i.e.
\begin{equation}
 {\rm WF}(u) := \bigcup_{a = 1}^r {\rm WF}(u_a),
\end{equation}
cf.\ \cite{Den}.\footnote{We assume that the reader is familiar with
  the  concept of the wavefront set of a scalar distribution, which is
  presented e.g.\ in the textbooks \cite{Hor1,Dieu7,RS2}.}
It is not difficult to check that this definition is, in fact,
independent of the choice of local trivialization of $\fX$ over $U$,
and thus yields a definition of WF$(u)$ for all $u \in (\Coin(\fX))'$
having the properties familiar of the wavefront set of scalar
distributions on the base manifold $N$, so that WF$(u)$ is a conical
subset of ${\rm T}^*N\backslash\{0\}$.

Another characterization of WF$(u)$ may be given in the following
way. Let $q \in U$ and $\xi \in {\rm T}^*_q N\backslash\{0\}$. Choose any
chart for $U$ around $q$, thus identifying $q$ with $0\in \bR^n$ and
$\xi$ with $\xi \in {\rm T}_0^*\bR^n \equiv \bR^n$ via the dual tangent
map of the chart. With respect to the chosen coordinates, we
introduce\\[4pt]
${}$\quad \quad \begin{tabular}{ll}
  translations: & $\check{\r}_x(y) := y + x$, \quad  and\\[2pt]
 dilations: &
   $\check{\d}_{\l}(y) := \l y$
\end{tabular}\\[4pt]
on a sufficiently small coordinate ball around $y = 0$ and taking $\l
> 0$ and the norm of $x \in \bR^n$ small enough so that the coordinate
range isn't left. Via pulling these actions back with help of the
chart they induce families of local diffeomorphisms $(\r_x)_{x \in B}$
and $(\d_{\l})_{0 < \l < \l_0}$ of $U$ for sufficiently small index ranges.
\\[2pt]
Now let ${\bf F}_q(\fX)$ be the set of all families $(f_{\l})_{\l > 0}$ of
sections in $\fX$ with
\begin{itemize}
\item[(i)] $f_{\l} \in \Coin(\fX_{\d_{\l}K})$ for some open
neighbourhood $K$ of $q$ when $\l$ is sufficiently small
\item[(ii)] For each continuous seminorm $\s$ on $\Coin(\fX)$
there is $s \ge 0$ so that $\sup_{\l}\,\l^s\s(f_{\l}) < \infty$.
\end{itemize}
With these conventions, we can formulate:
\begin{Lemma}
\label{lem2}
$(q,\xi)$ is not contained in {\rm WF}$(u)$ if and only if the
following holds:
\\[4pt]
For any family $(R_x)_{x \in B}$ of local morphisms of $\fX$ covering
$(\r_x)_{x \in B}$ there is some $h \in \cD(\bR^n)$ with $h(0) = 1$,
and an open neighbourhood $V$ of $\xi$ (in $\bR^n \equiv {\rm T}_q^*N$), such
that for all $(f_{\l})_{\l > 0} \in {\bf F}_q(\fX)$ one has
\begin{equation}
\label{equ15}
\sup_{k \in V}\, \left| \int {\rm e}^{-i\l^{-1}k\cdot
    x}h(x)\,u(R_x^{\stern} f_{\l})\,dx \right| = O^{\infty}(\l) \quad
{\rm as}\ \, \l \to 0\,.
\end{equation}
\end{Lemma}
\begin{proof}
Select a local trivialization of $\fX$ over $U$. With respect to it,
there are smooth GL$(r)$-valued functions $(R^a_b(x))_{a,b =1}^r$ of
$x$ such that\,\footnote{Summation over repeated indices will be assumed from
  now on. See also footnote 5.}
$$  u(R_x^{\stern}f_{\l}) = R^a_b(x)\,u_a(f_{\l}^b \lcrc \r_x^{-1})\,.
    $$
Now suppose that $(q,\xi)$ is not in WF$(u)$, so that $(q,\xi)$ isn't
contained in any of the WF$(u_a)$. Then, making use of the fact that
 the wavefront set of a scalar distribution may be
characterized in terms of the decay properties of its localized
Fourier-transforms in any coordinate chart (cf.\ \cite{Hor1}) in
combination with Prop. 2.1 and Lemma 2.2 in \cite{V.acs}, one obtains
immediately the relation \eqref{equ15}. Conversely, assume that \eqref{equ15}
holds. Since
$(R^a_b(x))_{a,b =1}^r$ is in GL$(r)$ for each $x$ and depends
smoothly on $x$, we can find a smooth family 
$(S^b_c(x))_{b,c =1}^r$ of functions of $x$ so that $S_c^b(x)R_b^a(x) =
\d_c^a$, $x \in B$. Since \eqref{equ15} holds, one may apply Lemma 2.2 of
\cite{V.acs} to the effect that for some open neighbourhood $V'$ of
$\xi$ and for all $(\,(0,\ldots,\varphi_{\l},\ldots,0)\,)_{\l > 0} \in
{\bf F}_q(\fX)$ where only the $c$-th entry is non-vanishing, one has
\begin{eqnarray*}
\lefteqn{\sup_{k \in V'}\,\left|\int {\rm e}^{-i\l^{-1}k\cdot
    x}h(x)\,u_c(\varphi_{\l}\lcrc\r_x^{-1})\,dx \right|}\\
& = & \sup_{k \in V'}\,\left|\int {\rm e}^{-i\l^{-1}k\cdot
    x}h(x)\,S_c^b(x)R_b^a(x)\,u_a(\varphi_{\l}\lcrc\r_x^{-1})\,dx
\right|\\
& = & O^{\infty}(\l) \quad {\rm as} \ \l \to 0\,.
\end{eqnarray*}
Then one concludes from Prop. 2.1 in \cite{V.acs} that $(q,\xi)$
isn't contained in WF$(u_c)$ for each
$c = 1,\ldots,r$.
\end{proof}
A very useful property is the behaviour of the wavefront set under
(local) morphisms of $\fX$. We put on record here the following Lemma
without proof, which may be obtained by extending the proof for the
scalar case in \cite{Hor1} together with some of the arguments appearing in the
proof of Lemma 3.1.
\begin{Lemma} 
\label{lem3}
Let $U_1$ and $U_2$ be open subsets of $N$, and let $R :
  \fX_{U_1} \to \fX_{U_2}$ be a vector bundle map covering a
  diffeomorphism $\r: U_1 \to U_2$. Let $u \in (\Coin(\fX_{U_1}))'$.
 Then it holds that
\begin{equation}
\label{equ3-trafo}
 {\rm WF}(R^{\stern}u) \subset {}^t D\r^{-1}{\rm WF}(u)\,,
\end{equation}
where ${}^t D\r^{-1}$ denotes the transpose (or dual) of the tangent
map of $\r^{-1}$. If $R$ is even a bundle morphism, then the inclusion
\eqref{equ3-trafo} specializes to an equality.
\end{Lemma}
\subsection{Briefing on Spacetime Geometry}
Since several concepts of spacetime geometry are going to play some
role lateron, we take the opportunity to introduce them here  and
establish the corresponding notation. We refer to the standard
references \cite{WaldI,HE} for a more thorough discussion and also
for definition of some well-established terminology that is not always
introduced explicitly in the following.

Let us assume that $(M,g)$ is a spacetime, so that $M$ is a smooth
manifold of dimension $m \ge 2$, and $g$ is a Lorentzian metric having
signature $(+,-,\ldots,-)$. It will also be assumed that the spacetime
is time-orientable, and that a time-orientation has been chosen. Then
one introduces, for any subset $G$ of $M$, the corresponding
future/past sets $J^{\pm}(G)$, consisting of all points lying on
piecewise smooth, continuous future/past-directed causal curves
emanating from $G$. A subset $G' \subset M$ is, by definition,
{\it causally separated} from $G$ if it has void intersection with
$\overline{J^+(G)} \cup \overline{J^-(G)}$. Thus a pair of points
$(q,p) \in M \times M$ is called causally separated if $q$ is causally
separated from $p$ or vice versa, since this relation is symmetric.

A smooth hypersurface $\Sigma$ in $M$ is called a {\it Cauchy-surface} if
each inextendible causal curve in $(M,g)$ intersects $\Sigma$ exactly
once. Spacetimes $(M,g)$ possessing Cauchy-surfaces are called  {\it
  globally hyperbolic}. It can be shown that a globally hyperbolic
spacetime admits smooth one-parametric foliations into Cauchy-surfaces.

Globally hyperbolic spacetimes have a very well-behaved causal
structure. A certain property of globally hyperbolic spacetimes will
be important for applying the propagation of singularities theorem
in Section 5, so we mention it here: Let $v$ be a
non-zero lightlike vector in ${\rm T}_qM$ for some $q \in M$.
It defines a maximal
smooth, affinely parametrized geodesic $\g : I \to M$ with the
properties $\g(0) = q$ and $\left.\frac{d}{dt}\g(t)\right|_{t = 0} =
v$ where `maximal' here refers to choosing $I$ as the largest real
interval ($I$ is taken as a neighbourhood of 0, and may coincide e.g.\
with $\bR$) where $\g$ is a smooth solution of the geodesic equation
compatible with the specified data at $q$. Then $\g$ is both future-
and past-inextendible 
(see e.g.\ the argument in \cite[Prop.\ 4.3]{Rad1}), and 
consequently, given an arbitrary
Cauchy-surface $\Sigma \subset M$, there is exactly one parameter value $t
\in I$ so that $\g(t) \in \Sigma$. 
\subsection{Wave-Operators and Propagation of Singularities}
%
Suppose that we are given a time-oriented spacetime $(M,g)$. Then let $\fV$ be
a vector bundle with base manifold $M$, typical fibre isomorphic to
$\bC^{\,r}$, and bundle projection $\p_M$. Moreover, we assume that
there exists a morphism $\G$ of $\fV$ covering the identity map of $M$
which is involutive ($\G \lcrc \G = {\rm id}_{\fV}$) and acts
anti-isomorphically on the fibres; in other words, $\G$ acts like a
complex conjugation in each fibre space. Therefore, the $\G$-invariant
part $\fV^{\circ}$ of $\fV$ is a vector bundle over the base $M$ with
typical fibre $\bR^{\,r}$.

A linear partial differential operator
$$ P : \Coin(\fV) \to \Coin(\fV) $$
will be said to have {\it metric principal part} if, upon choosing a
local trivialization of $\fV$ over $U \subset M$ in which sections $f
\in \Coin(\fV_U)$ take the component representation
$(f^1,\ldots,f^r)$, and a chart $(x^{\m})_{\m =1}^m$, one has the
following coordinate representation for $P$:\,\footnote{Greek indices
  are raised and lowered with $g^{\m}{}_{\n}(x)$, latin indices with $\d^a_b$.}
$$ (Pf)^a(x) = g^{\m\n}(x)\partial_{\m}\partial_{\n}f^a(x) +
A^{\n}{}^a_b(x)\partial_{\n}f^b(x) + B^a_b(x)f^b(x)\,.$$
Here, $\partial_{\m}$ denotes the coordinate derivative
$\frac{\partial}{\partial x^{\m}}$, and $A^{\n}{}^a_b$ and $B^a_b$ are
suitable collections of smooth, complex-valued functions. Observe that
thus the principal part of $P$ diagonalizes in all local
trivializations (it is ``scalar'').

If $P$ has metric principal part and is in addition $\G$-invariant,
i.e.\
\begin{equation}
 \G^{\stern} \lcrc P \lcrc \G^{\stern} = P\,,
\end{equation}
then we call $P$ a {\it wave operator}.
In this case, $P$ leaves the space $\Coin(\fV^{\circ})$ of
$\G^{\stern}$-invariant 
sections invariant. As an aside we note that then there is a covariant
derivative (linear connection) $\nabla^{(P)}$ on $\fV^{\circ}$
together with a bundle map $v$ of $\fV^{\circ}$ covering ${\rm id}_M$
such that
$$ Pf = g^{\m\n}\nabla^{(P)}_{\m}\nabla^{(P)}_{\n}f + v^{\stern}f $$
for all $f \in \Coin(\fV^{\circ})$; this covariant derivative is given
by
$$ 2\cdot \nabla_{{\rm grad}\varphi}^{(P)}f = P(\varphi f) - \varphi
P(f) - (\Box_g\varphi)f $$
for all $\varphi \in \Coin(M,\bR)$ and $f \in \Coin(\fV^{\circ})$,
where $\Box_g$ denotes the d'Alembert-operator induced by the metric
$g$ on scalar functions \cite{Gun}.

Before we can state the version of the propagation of singularities
theorem that will be relevant for our considerations later,
 we need to introduce further notation. By $\fV \bt \fV$ we
denote the outer product bundle of $\fV$. This is the $\Cin$-vector
bundle over $M \times M$ whose fibres over $(q_1,q_2) \in M \times M$
are $\fV_{q_1} \otimes \fV_{q_2}$ where $\fV_{q_j}$ denotes the fibre
over $q_j$ ($j = 1,2$), and with base projection defined by
$$ {\rm v}_{q_1} \otimes {\rm v'}_{q_2} \mapsto (q_1,q_2) \quad {\rm
  for} \quad {\rm v}_{q_1}\otimes {\rm v'}_{q_2} \in \fV_{q_1} \otimes
\fV_{q_2}\,.$$
Note also that the conjugation $\G$ on $\fV$ induces a conjugation $\bt^2\G$
on $\fV \bt \fV$ by anti-linear extension of the assignment
$$ \bt^2\G({\rm v}_{q_1} \otimes {\rm v'}_{q_2}) := \G{\rm v}_{q_1}
\otimes \G{\rm v'}_{q_2}\,, \quad q_j \in M\,.$$
The definition of $\bt^n\fV$, the $n$-fold outer tensor product of
$\fV$, should then be obvious, and likewise the definition of
$\bt^n\G$.

Going to local trivializations and using partition of unity arguments,
it is not difficult to see that the canonical embedding $\Coin(\fV)
\otimes \Coin(\fV) \subset \Coin(\fV \bt \fV)$ is dense (\cite{Dieu7}).
Moreover, if we take some $L \in (\Coin(\fV \bt \fV))'$, then it
induces a bilinear form $\L$ over $\Coin(\fV)$ by setting
\begin{equation}
\label{equ16}
 \L(f,f') = L(f \otimes f')\,,\quad f,f' \in \Coin(\fV)\,.
\end{equation}
Clearly $\L$ is then jointly continuous in both entries. On the other
hand, if $\L$ is a bilinear form over $\Coin(\fV)$ which is separately
continuous in both entries ($f \mapsto \L(f,f')$ and $f \mapsto
\L(f',f)$ are continuous maps for each fixed $f'$), then the nuclear
theorem implies that there is an $L \in (\Coin(\fV \bt \fV))'$ inducing
$\L$ according to \eqref{equ16} \cite{Dieu7}. These statements generalize to
the case of $n$-fold tensor products in the obvious manner.

Now define \footnote{The notation $(q,\xi) \in {\rm T}^*M$ means that
  $\xi \in {\rm T}^*_qM$, i.e.\ $q$ denotes the base point of the
  cotangent \\  vector $\xi$.}
$$ \cN := \{(q,\xi) \in {\rm T}^*M\backslash\{0\}:
g^{\m\n}(q)\xi_{\m}\xi_{\n} = 0\}\,.$$
Moreover, define for each pair $(q,\xi;q',\xi') \in \cN \times \cN$:
\\[4pt]
${}$ \quad \quad $(q,\xi) \sim (q',\xi')$ \quad iff there exists
an affine parametrized lightlike geodesic $\g$ in $(M,g)$ connecting
$q$ and $q'$ and such that $\xi$ and $\xi'$ are co-tangent to
$\g$ at $q$ and $q'$, respectively.
\\[4pt]
Here, we say that $\xi$ is co-tangent to $\g$ at $q = \g(s)$ if
$(\left.\frac{d}{dt}\right|_{t = s}\g(t))^{\m} =
g^{\m\n}(q)\xi_{\n}$, where $t$ is the affine
parameter. Therefore, $(q,\xi) \sim (q',\xi')$ means
 $\xi$ and $\xi'$ are parallel transports of
each other along the lightlike geodesic $\g$ connecting $q$ and $q'$.
Note that the possibility $q = q'$ is included, in
which case $(q,\xi) \sim (q',\xi')$ means $\xi = \xi'$. 
One can introduce the following two disjoint future/past-oriented parts
(with respect to the time-orientation of $(M,g)$) of $\cN$, 
\begin{equation}
\label{equ16a}
\cN_{\pm}:=\{(q,\xi)\in\cN\;\vert\;\pm\xi\bef 0 \}\,,
\end{equation}
where $\xi \bef 0$ means that the vector $\xi{}^{\m} =
g^{\m\n}\xi_{\n}$ is future-pointing.

The relation ``$\sim$'' is obviously an equivalence relation between
elements in $\cN$. For $(q,\xi) \in \cN$, the corresponding
equivalence class is denoted by B$(q,\xi)$; it is a bi-characteristic
strip of any wave operator $P$ on $\fV$ since such an operator has
metric principal part and therefore its bi-characteristics are
lightlike geodesics (see, e.g.\ \cite{KRW}).

Now we are ready to state a specialized version of the propagation of
singularities theorem (PST) which is tailored for two-point
distributions that are solutions (up to $C^{\infty}$-terms) of wave
operators, and which derives as a special case of the PST in
\cite{Den}.
We should like to point out that the formulation of the PST in
\cite{Den} (extending arguments developed in \cite{DH} for the scalar
case) is considerably more general in two respects: First, it applies,
with suitable modifications, not only to linear second order
differential operators with metric principal part, but to
pseudo-differential operators on $\Coin(\fV)$ that have a so-called
`real principal part' (of which `metric principal part' is a special
case, note also that a metric principal part is homogeneous).
 Secondly, the general formulation of the PST  gives not only
information about the wavefront set of a $u \in (\Coin(\fV))'$ which
is a solution up to $C^{\infty}$-terms of a pseudo-differential
operator $A$ having real principal part (i.e.\ WF$(Au) = \emptyset$),
but even describes properties of the polarization set of such a $u$.
The polarization set WF$_{\rm pol}(u)$ of $u \in (\Coin(\fV))'$ is a
subset of the direct product bundle ${\rm T}^*M \oplus \fV$ over $M$ and
specifies which components of $u$ (in a local trivialization of $\fV$)
have the worst decay properties in Fourier-space near any given base
point in $M$; the projection of WF$_{\rm pol}(u)$ onto its ${\rm T}^*M$-part
coincides with the wavefront set WF$(u)$. The reader is referred to
\cite{Den} for details and further discussion, and also to \cite{Krat,Hol}
for a discussion of the polarization set for Dirac fields on curved
spacetimes.  As a corollary to the
PST formulated in \cite{Den} together with Lemma 6.5.5.\ in
\cite{DH} (see also \cite{KRW} for an elementary account),
 one obtains the following:
\begin{Prop}
\label{pro2}
Let  $P$ be a wave operator on $\Coin(\fV)$ and define
for $w \in (\Coin(\fV \bt \fV))'$ the distributions
$w_{(P)},w^{(P)} \in (\Coin(\fV \bt \fV))'$ by
\begin{equation}
\label{equ18}
\begin{split}
 w_{(P)}(f \otimes f') & := w(Pf\otimes f')\,,\\[2pt]
 w^{(P)}(f \otimes f') & := w(f \otimes Pf')\,,
\end{split}
\end{equation}
for all $f,f' \in \Coin(\fV)$.

Suppose that ${\rm WF}(w_{(P)}) = \emptyset = {\rm
  WF}(w^{(P)})$. Then it holds that
$$  {\rm WF}(w) \subset \cN \times \cN $$
and
$$ (q,\xi;q',\xi') \in {\rm WF}(w)\ \ {\rm with}\ \ \xi \ne 0 \ {\rm
  and}\ \xi' \ne 0 \ \  \Rightarrow \ \  
 {\rm B}(q,\xi) \times {\rm B}(q',\xi') \subset {\rm WF}(w)\,.$$
\end{Prop}

\section{Quantum Fields}
\setcounter{equation}{0}
\label{sec1}
\subsection{The Borchers Algebra}
\label{sec4}
We begin our discussion of linear quantum fields obeying a wave
equation by recalling the definition and basic properties of the
Borchers-algebra \cite{Bor}.

Let $\fV$ denote a vector bundle over the base-manifold $M$ as in the
previous section. Then consider the set
$$
   \cB:=\{\fv \equiv (\fv_n)_{n = 0}^{\infty}: \fv_0 \in \bC,\ \fv_n \in
   \Coin(\bt^n\fV), \ {\rm only\ finitely\ many}\ \fv_n \ne 0\} $$
where $\bt^n\fV$ denotes the $n$-fold outer product bundle of $\fV$,
cf.\ Sec.\ \ref{sec2}. The set $\cB$ is a priori a vector space, but one may
also introduce a $*$-algebraic structure on it: A product $\fv \cdot
\gv$ for elements $\fv, \gv \in \cB$ is given by defining the $n$-th
component $(\fv \cdot \gv)_n$ to be
$$ (\fv \cdot \gv)_n := \sum_{i + j = n}\fv_i \otimes \gv_j\,.$$
Here, $\fv_i \otimes \gv_j$ is understood as the element in
$\Coin(\bt^n\fV)$ induced by the canonical embedding
$\Coin(\bt^i\fV)\otimes \Coin(\bt^j\fV) \subset
\Coin(\bt^n\fV)$. Observe that $\cB$ possesses a unit element
$1_{\cB}$, given by the sequence $((1_{\cB})_n)_{n = 0}^{\infty}$
having the number $1$ in the 0-th component while all other components
vanish. Moreover, for $\fv \in \cB$ one can define $\fv^*$ by setting
\begin{equation}
\label{equ1}
\fv^*_n(q_1,\ldots,q_n) :=\bt^n \G\,\fv_n(q_n,\ldots,q_1)\,, \quad q_j \in
M\,,
\end{equation}
for the $n$-th component of $\fv^*$ where $\G$ denotes the complex
conjugation assumed to be given on $\fV$. This yields an anti-linear
involution on $\cB$. With these definitions of product and
$*$-operation, $\cB$ is a $*$-algebra.

Furthermore, $\cB$ has a natural `local net structure'
in the sense that one obtains an inclusion-preserving map $M \supset \cO
\mapsto \cB(\cO) \subset \cB$ taking subsets $\cO$ of $M$ to
unital $*$-subalgebras $\cB(\cO)$ of $\cB$ upon defining $\cB(\cO)$ to
consist of all $(\fv_n)_{n = 0}^{\infty}$ for which ${\rm supp}\,\fv_n
\subset \cO$, $n \in \bN$.

Another simple fact is that (local) morphisms of $\fV$
commuting with $\G$ can be lifted
to (local) automorphisms of $\cB$. To this end, let $(R_x)_{x \in B}$
be a family of (local) morphisms of $\fV$ covering $(\r_x)_{x \in
  B}$, and assume that $\G R_x = R_x\G$ for all $x$.
 Suppose that $\cO \subset M$ is in the domain of $\r_x$; then
define a map $\a_x$ on $\cB(\cO)$ by setting for $\fv \in \cB(\cO)$ the $n$-th
component, $(\a_x \fv)_{n}$, of $\a_x\fv$ to be
\begin{equation}
\label{equ1a}
(\a_x\fv)_n := \bt^nR_x^{\stern}\,\fv_n\,,
\end{equation}
where
$$ \bt^nR^{\stern}_x(g^{(1)}\otimes \cdots \otimes g^{(n)}) :=
R_x^{\stern}g^{(1)}\otimes \cdots \otimes R_x^{\stern}g^{(n)}\,, \quad
g^{(j)} \in \Coin(\fV)\,,$$
defines the outer product action of $R_x^{\stern}$ via linear
extension on $\Coin(\bt^n\fV)$. It is not difficult to check that this
yields a $*$-isomorphism $\a_x:\cB(\cO) \to \cB(\r_x(\cO))$.

We will now turn $\cB$ into a locally convex space by 
giving it the topology of the 
strict inductive limit of the toplogical vector spaces
\begin{equation*}
\cB_n:=\bC\oplus\bigoplus_{k=1}^n \Coin(\bt^k\fV),\quad\quad\quad n\in\bN.
\end{equation*}
This topology is known as the \textit{locally convex direct sum topology} 
(ex. \cite[Chap. II, \S 4 n$^\circ$ 5]{Bou}).
Some important properties of $\cB$, equipped with this topology are
given in the following lemma, the proof of which will be deferred to
Appendix \ref{app2}.
\begin{Lemma}
\label{lem1}
With the topology given above, $\cB$ is complete and a topological 
$*$-algebra.  
Moreover, a linear functional $u:\cB \to \bC$ is continuous if and
only if there is a sequence $(u_n)_{n = 0}^{\infty}$ with $u_0 \in
\bC$ and $u_j \in (\Coin(\bt^j\fV))'$ for $j \in \bN$ so that
\begin{equation}
\label{equ1d} 
u(\fv) = u_0\fv_0 + \sum_{j\in \bN}u_j(\fv_j)\,,\quad \fv \in
\cB\,.
\end{equation}
If $\alpha$ is a $*$-automorphism lifting a morphism 
$R$ of $\fV$ to $\cB$ as in \eqref{equ1a}, then $\alpha$ is
continuous.
Moreover, let $(R_x)_{x \in B}$ be a family of morphisms of $\fV$
depending smoothly on $x$ with $\G R_x = R_x \G$ and $R_0 = {\rm
  id}_{\fV}$, and let $(\a_x)_{x \in B}$ be the 
family of $*$-automorphisms of $\cB$ induced according to
 \eqref{equ1a}.  Then for
each ${\sf f} \in \cB$ it holds that 
\begin{equation}
\label{equ1b}
\alpha_x(\fv)\to \fv \quad\text{ for }
\quad x\to 0,
\end{equation}
and there is a constant $r>0$ such that to each continuous semi-norm 
$\sigma$ of $\cB$ one can find another semi-norm $\sigma'$ with the property
\begin{equation}
\label{equ1c}
\sigma(\alpha_x(\fv))\leq\sigma'(\fv)\,, \quad |x| \le r,\ {\sf f} \in \cB\,.
\end{equation}
\end{Lemma}
\subsection{States and Quantum Fields}
A state $\o$ on $\cB$ is a continuous linear form on $\cB$  which fulfills the
positivity requirement $\o(\fv^*\fv) \ge 0$ for all $\fv \in \cB$.
By Lemma \ref{lem1} such a state $\omega$ is completely characterized by 
a set $\{\omega_n\vert n\in\NN_0\}$ of linear functionals 
$\omega_n\in (\Coin(\bt^n\fV))'$, the so-called \textit{$n$-point
functions}.

The positivity requirement
allows it to associate with any state $\o$ a Hilbertspace
$*$-representation by the well-known Gelfand-Naimark-Segal (GNS)
construction (or the Wightman reconstruction theorem \cite{SW}). More
precisely, given a state on $\cB$, there exists a triple $(\varphi,\cD
\subset \cH, \O)$, called GNS-representation of $\o$, possessing the
following properties:
\begin{itemize}
\item[(a)] $\cH$ is a Hilbertspace, and  $\cD$ is a dense linear
  subspace of $\cH$.
\item[(b)] $\varphi$ is a $*$-representation of $\cB$ on $\cH$ by closable
  operators with common domain $\cD$.
\item[(c)] $\O$ is a unit vector contained in $\cD$ which is cyclic,
  i.e.\ $\cD = \varphi(\cB)\O$, and has the property that
$$ \o(\fv) = \langle  \O,\varphi(\fv)\O\rangle \, ,\quad \fv \in
\cB\,.$$
\end{itemize}
Furthermore, the GNS-representation is unique up to unitary equivalence.
We refer to \cite[Part II]{Sch} for further details on $*$-representations of
$*$-algebras as well as for a proof of these statements and references
to the relevant original literature.

Therefore, a state $\o$ on $\cB$ induces a quantum field --- that is
to say, an operator-valued distribution
\begin{equation}
\label{equ17}
\Coin(\fV) \owns f  \mapsto \Phi(f) := \varphi(\fv)\,, \quad \fv =
(0,f,0,0,\dots )\,,
\end{equation}
where the $\Phi(f)$ are, for each $f \in \Coin(\fV)$, closable
operators on the dense and invariant domain $\cD$ and one has
$\Phi(\G^{\stern}f) \subset \Phi(f)^*$ where $\Phi(f)^*$ denotes the
adjoint operator of $\Phi(f)$. Conversely, such a quantum field
induces states on $\cB$: Given some unit vector $\psi \in \cD$, the
assignment
\begin{eqnarray*}
 \o^{(\psi)}(c\cdot 1_{\cB}) & := & c\,, \quad c \in \bC\,,\\
 \o^{(\psi)}(f^{(1)}\otimes \cdots \otimes f^{(n)}) & := & \langle
 \psi,\Phi(f^{(1)})\cdots \Phi(f^{(n)})\psi\rangle
\,, \quad f^{(j)} \in \Coin(\fV)\,,
\end{eqnarray*}
defines, by linear extension, a state $\o^{(\psi)}$ on $\cB$. (Obviously this
generalizes from vector states to mixed states.)

If the quantum field $\Phi$ is an observable field, then one would
require commutativity at causal separation, and this means
$$ \Phi(f)\Phi(f') = \Phi(f')\Phi(f) $$
whenever the supports of $f$ and $f'$ are causally separated.
 Such commutative behaviour (locality) of $\Phi$ at causal
separation is characteristic of bosonic fields. On the other hand, a
field $\Phi$ is fermionic if it anti-commutes at causal separation
(twisted locality), i.e.\
$$ \Phi(f)\Phi(f') = - \Phi(f')\Phi(f) $$
for causally separated supports of $f$ and $f'$.
The general analysis of quantum field theory so far has shown that the
alternative of having quantum fields of bosonic or fermionic character
may largely be viewed as generic at least for spacetime dimensions
greater than 2 \cite{Haag,SW,DR,GLRV}.

 If $\o$ is a state on $\cB$ inducing via its GNS-representation a
 bosonic field, then it follows that the commutator $\o^{(-)}_2$ of
 its two-point function, defined by 
$$ \o^{(-)}_2(f \otimes f') := \frac{1}{2}(\o_2(f \otimes f') -
\o_2(f'\otimes f))\,, \quad f,f' \in \Coin(\fV)\,,$$
vanishes as soon as the supports of $f$ and $f'$ are causally
separated. If, on the other hand, $\o$ induces a fermionic field, then
the anti-commutator,
$$ \o^{(+)}_2(f \otimes f') := \frac{1}{2}(\o_2(f \otimes f') +
\o_2(f'\otimes f))\,, \quad f,f' \in \Coin(\fV)\,,$$
of its two-point function vanishes when the supports of $f$ and $f'$
are causally separated.

For our purposes in Sec.\ \ref{sec3}, we may assume a weaker version of
bosonic or fermionic behaviour of quantum fields: We shall later
suppose that $\o^{(+)}_2$ or $\o^{(-)}_2$ is smooth ($C^{\infty}$) at
causal separation. The definition relevant for that terminology is as
follows:
\begin{Dfn} 
\label{def1}
  Let $w \in (\Coin(\fV \bt \fV))'$. We say that $w$ is
  {\it smooth at causal separation} if 
 $$ {\rm WF}(w_{\cQ}) = \emptyset $$
where $\cQ$ is the set of all pairs of points $(q,q') \in M \times M$
which are causally separated in $(M,g)$ \footnote{\rm  $\cQ$ is an
  open subset in $M \times M$ due to global hyperbolicity.}
 and $w_{\cQ}$ denotes the restriction of $w$ to $\Coin((\fV \bt \fV)_{\cQ})$.
\end{Dfn}
\subsection{Quasifree States}
Of particular interest are quasifree states associated with quantum
fields obeying canonical commutation relations (CCR) or canonical
anti-commutation relations (CAR). A simple way of introducing them is
via the characterization of such states given in \cite{Kay.qf} which
we will basically follow here. Note, however, that in this reference
the map $K$  in \eqref{equ.qf} is defined on certain quotients
of $\Coin(\fV^{\circ})$ while we define $K$ on $\Coin(\fV^{\circ})$
itself (recall that $\Coin(\fV^{\circ})$ is the space of
$\G^{\stern}$-invariant sections).
This is due to the fact that we haven't imposed CCR or CAR for states
on the Borchers algebra, so the notion of quasifree states given here
is, in this respect, more general.

Let $\hh$ be a complex Hilbertspace (the so called `one-particle
Hilbertspace') and $F_{\pm}(\hh)$ the bosonic/fermionic Fock-space
over $\hh$. By $a_{\pm}(\,.\,)$ and $a_{\pm}^{\dagger}(\,.\,)$ we
denote the corresponding annihilation and creation operators,
respectively. The Fock-vacuum vector will be denoted by $\O_{\pm}$.
Then we say that a state $\o$ on $\cB$ is a {\it (bosonic/fermionic)
  quasifree state} if there exists a real-linear map
\begin{equation}
\label{equ.qf}
 K : \Coin(\fV^{\circ}) \to \hh
\end{equation} 
whose complexified range is dense in $\hh$, such that the
GNS-representation $(\varphi,\cD \subset\cH,\O)$ of $\o$ takes the
following form: $\cH = F_{\pm}(\hh)$, $\O = \O_{\pm}$, and
$$ \Phi(f) = \frac{1}{\sqrt{2}}\left( a_{\pm}(K(f)) +
  a^{\dagger}_{\pm}(K(f))\right) \,,\quad f \in
\Coin(\fV^{\circ})\,,$$
where $\Phi(\,.\,)$ relates to $\varphi(\,.\,)$ as in \eqref{equ17}.

Quasifree states are in a sense the most simple
states. It is, however, justified to consider prominently those
states since for quantum fields obeying a linear wave-equation,
ground- and KMS-states turn out to be quasifree in examples.
Any quasifree state $\o$ is entirely determined by its two-point
function, i.e.\ by the map
$$ \Coin(\fV) \times \Coin(\fV) \owns (f^{(1)},f^{(2)}) \mapsto
\omega(f^{(1)} \otimes f^{(2)}) = \langle
\O,\Phi(f^{(1)})\Phi(f^{(2)})\O\rangle\,, $$
in the sense that the $n$-point functions 
$$ \o_n(f^{(1)}\otimes \cdots \otimes f^{(n)}) = \langle
\O,\Phi(f^{(1)})\cdots \Phi(f^{(n)})\O \rangle\,, \quad f^{(j)} \in
\Coin(\fV)\,,$$
vanish for all odd $n$, while the $n$-point functions for even $n$ can
be expressed as polynomials in the variables $\o_2(f^{(i)} \otimes
f^{(j)})$, $i,j = 1,\ldots,n$. This attaches
particular significance to the two-point functions for
quantum fields obeying linear wave equations. We refer to
\cite{BR2,Kay.qf,Araki} for further discussion of quasifree states and
their basic properties.
\section{Passivity and Microlocal Spectrum Condition}
\setcounter{equation}{0}
\label{sec3}
In the present section we will state and prove our main result
connecting passivity and microlocal spectrum condition for linear
quantum fields obeying a hyperbolic wave equation on a 
globally hyperbolic, stationary spacetime.

First, we need to collect the assumptions. It will be assumed that
$\fV$ is a vector bundle, equipped with a conjugation $\G$, over a base
manifold $M$ carrying a time-orientable Lorentzian metric
$g$, and that $(M,g)$ is globally hyperbolic.  Moreover, we assume
that the spacetime $(M,g)$ is stationary, so
that there is a one-parametric $C^{\infty}$-group $\{\t_t\}_{t\in\bR}$
of isometries whose generating vector field, denoted by
$\partial^{\t}$, is everywhere timelike and future-pointing
(with respect to a fixed time-orientation). We recall that the
notation $\cN_{\pm}$ for the future/past-oriented parts of the set of
null-covectors $\cN$ has been introduced in \eqref{equ16a}, and note
that $(q,\xi) \in \cN_{\pm}$ iff $\pm \xi(\partial^{\t}) > 0$.
It is furthermore supposed that there is a smooth one-parametric group
$\{T_t\}_{t\in\bR}$ of morphisms of $\fV$ covering
$\{\t_t\}_{t\in\bR}$, and a wave operator $P$ on $\Coin(\fV)$, having
the following properties:
$$ \G \lcrc T_t = T_t \lcrc \G \,, \quad \ T^{\stern}_t \lcrc P = P
\lcrc T^{\stern}_t\,, \quad \ t\in\bR\,. $$
Now let $\cB$ again denote the Borchers algebra as in Sec.\ \ref{sec1}. The
automorphism group induced by lifting $\{T_t\}_{t\in\bR}$ on $\cB$
according to \eqref{equ1a} will be denoted by
$\{\a_t\}_{t\in\bR}$. Whence, by Lemma 4.1,
$(\cB,\{\at\}_{t\in\bR})$ is a topological $*$-dynamical system.

Recall that a state $\o$ on $\cB$ is, by definition, contained in
$\cP$ if it is a convex combination of ground- or KMS-states at
strictly positive inverse temperature for $\{\at\}_{t\in\bR}$.
\begin{Thm}
\label{the1}
Let $\o \in \cP$ and let $\o_2$ be the two-point distribution of $\o$ (see
Sec.\ \ref{sec4}). Suppose that ${\rm WF}(\o_2^{(P)}) = \emptyset = {\rm
  WF}(\o_2{}_{(P)})$ where $\o_2^{(P)}$ and $\o_2{}_{(P)}$ are defined
as in \eqref{equ18}, and suppose also that the symmetric part $\o_2^{(+)}$ or
the anti-symmetric part $\o^{(-)}_2$ of the two-point distribution
is smooth at causal separation (Definition \ref{def1}).

Then it holds that  ${\rm WF}(\o_2)\subset\cR$ where
$\cR$ is the set  
\begin{equation}
\label{equ26}
\cR := \{ (q,\xi;q',\xi') \in \cN_- \times \cN_+ : (q,\xi) \sim
(q',-\xi')\}.
\end{equation}

\end{Thm}
\begin{proof} {\it 1.) } Let $q$ be any point in $M$. Then there is a
  coordinate chart $\k = (y^0,\yu) = (y^0,y^1,\ldots,y^{m-1})$ around
  $q$ so that, for small $|t|$, 
$$ \k \lcrc \t_t = \check{\t}_t \lcrc \k $$
holds on a neighbourhood of $q$, where
$$ \check{\t}_t(y^0,\yu) := (y^0 + t,\yu)\,.$$
In such a coordinate system, we can also define ``spatial''
translations
$$ \check{\r}_{\xu}(y^0,\yu) := (y^0,\yu + \xu) $$
for $\xu = (x^1,\ldots,x^{m-1})$ in a sufficiently small neighbourhood
$\Bu$ of the origin in $\bR^{m-1}$. Let $(R_{\xu})_{\xu \in \Bu}$ be any
smooth family of local morphisms around $q$ covering $(\r_{\xu})_{\xu
  \in \Bu}$, where $\r_{\xu} := \k^{-1} \lcrc \check{\r}_{\xu} \lcrc
\k$ (on a sufficiently small neighbourhood of $q$). Now let $q'$ be
another point, and choose in an analogous manner as for $q$ a
coordinate system $\k'$, and $(\r'_{\xu'})_{\xu' \in \Bu'}$ and
$(R'_{\xu'})_{\xu' \in \Bu'}$.
\\[6pt]
{\it 2.) } In a further step we shall now establish the relation
\begin{equation}
\label{equ23}
{\rm WF}(\o_2) \subset \{(q,\xi;q',\xi') \in ({\rm T}^*M \times {\rm
  T}^*M)\backslash \{0\} :
\xi(\partial^{\t}) + \xi'(\partial^{\t}) = 0, \ \, \xi'(\partial^{\t})
 \ge 0\}\,.
\end{equation}
Since we have ${\rm WF}(\o_2) \subset \cN \times \cN$ by Prop.\ 3.2,
this then allows us to conclude that 
\begin{equation}
\label{equ23a}
{\rm WF}(\o_2) \subset \{(q,\xi;q',\xi') \in \cN_- \times \cN_+ :
\xi(\partial^{\t}) + \xi'(\partial^{\t}) = 0\}\,,
\end{equation} 
and we observe that thereby the possibility $(q,\xi;q',\xi') \in {\rm
  WF}(\o_2)$ with $\xi = 0$ or $\xi' = 0$ is excluded, because that
would entail both $\xi = 0$ and $\xi' = 0$.

For proving \eqref{equ23} it is in view of Lemma \ref{lem2} and
 according to our choice
of the coordinate systems  $\k$, $\k'$ and corresponding actions
$(R_{\xu})_{\xu \in \Bu}$ and $(R'_{\xu'})_{\xu' \in \Bu'}$ sufficient
to demonstrate that the following holds:
\\[4pt]
There is a function $h \in \Coin(\bR^m \times \bR^m)$ with $h(0) = 1$,
and for each $(\xi;\xi') = (\xi_0,\xiu;\xi'_0,\xiu') \in (\bR^m \times
\bR^m)\backslash \{0\}$ with $\xi_0 + \xi_0' \ne 0$ or $\xi_0' < 0$
there is an open neighbourhood $V \subset (\bR^m \times \bR^m)\backslash
\{0\}$ so that 
{\small
\begin{equation}
\label{equ19}
\sup_{(k;k') \in V} \left|
\int {\rm e}^{-i\l^{-1}(tk_0 + \xu\cdot\ku)}{\rm e}^{-i\l^{-1}(t'k_0' +
  \xu' \cdot \ku')}h(t,\xu;t',\xu')\o_2((T^{\stern}_tR^{\stern}_{\xu}
\otimes T^{\stern}_{t'}R'_{\xu'}\!{}^{\stern})F_{\l})\,dt\,dt'\,d\xu\,d\xu'
 \right| = O^{\infty}(\l)
\end{equation}}
as $\l \to 0$ holds for all $(F_{\l})_{\l > 0} \in {\bf
  F}_{(q;q')}(\fV \bt \fV)$. (The notation $k = (k_0,\ku)$ should be
obvious.) However, making use of part (c) of the statement of Prop.\ 
2.1 in \cite{V.acs}, for proving \eqref{equ19} it is actually enough to show
that there are $h$ and $V$ as above so that
{\small
\begin{equation}
\label{equ22}
\sup_{(k;k') \in V} \,\left|
\int {\rm e}^{-i\l^{-1}(tk_0 + \xu\cdot\ku)}{\rm e}^{-i\l^{-1}(t'k_0' +
  \xu' \cdot \ku')}h(t,\xu;t',\xu')\o_2(T^{\stern}_tR^{\stern}_{\xu} f_{\l}
\otimes T^{\stern}_{t'}R'_{\xu'}\!{}^{\stern}f'_{\l})\,dt\,dt'\,d\xu\,d\xu'
 \right| = O^{\infty}(\l)
\end{equation}}
as $\l \to 0$ holds for all $(f_{\l})_{\l > 0} \in {\bf F}_{q}(\fV)$
and all $(f'_{\l})_{\l > 0} \in {\bf F}_{q'}(\fV)$.

In order now to exploit the strict passivity of $\o$ via
Prop. \ref{pro1}, we define
the set ${\bf B}^{\sharp}$ of testing families with respect to the
Borchers algebra $\cB$ in the same  manner as we have defined the set ${\bf
  A}^{\sharp}$ of testing families for the algebra $\cA$ in Remark
\ref{rem1}. In other words, a $\cB$-valued family $({\sf f}_{z,\l})_{\l >
  0,z\in\bR^n}$ is a member of ${\bf B}^{\sharp}$, for arbitrary $n
\in \bN$, whenever for each continuous seminorm $\s$ on $\cB$ there is
some $s \ge 0$ so that 
$$ \sup_{z,\l}\,\l^s\s({\sf f}_{z,\l}^*{\sf f}_{z,\l}) < \infty\,. $$
Now if $(f_{\l})_{\l > 0}$ is in ${\bf F}_q(\fV)$, then 
$({\sf f}_{\xu,\l})_{\l > 0,\xu \in \Bu}$ defined by
\begin{equation}
\label{equ20}
{\sf f}_{\xu,\l} := (0,R_{\xu}^{\stern}f_{\l},0,0,\ldots)
\end{equation}
is easily seen to be a testing family in ${\bf B}^{\sharp}$.
The same of course holds when taking any $(f'_{\l})_{\l > 0} \in {\bf
  F}_{q'}(\fV)$ and defining $({\sf f}'_{\xu',\l})_{\l > 0,\xu' \in
  \Bu'}$ accordingly.

Since $\o \in \cP$, it follows from Prop. \ref{pro1} and Remark
\ref{rem1} that,
with respect to the time-translation group $\{\at\}_{t\in \bR}$, 
$${\it ACS}^2_{{\bf B}^{\sharp}}(\o) 
\subset \{(\xi_0,\xi_0') \in \bR^2\backslash\{0\} : \xi_0 + \xi_0' =
0,\ \xi'_0 \ge 0\} \,. $$
And this means that
there is some  $h_0 \in \Coin(\bR^2)$  with $h_0(0)=1$, and for each
$(\xi_0,\xi_0')\in \bR^2\backslash\{0\}$ with $\xi_0 + \xi_0' \ne 0$
or $\xi'_0 < 0$  an open neighbourhood $V_0$ in $\bR^2\backslash\{0\}$
so that
\begin{equation}
\label{equ21}
\sup_{(k_0,k_0') \in V_0,\,\xu,\xu'}\,
\left| \int {\rm e}^{-i\l^{-1}(t k_0 + t'k_0')}h_0(t,t')\o(\a_{t}({\sf
      f}_{\xu,\l}) \a_{t'}({\sf f}'_{\xu',\l}))\,dt\,dt' \right| =
  O^{\infty}(\l) 
\end{equation}
as $\l \to 0$ for all $({\sf f}_{\xu,\l})_{\l < 0,\xu \in \Bu}$ and
$({\sf f}'_{\xu',\l})_{\l > 0,\xu'\in\Bu'}$ in ${\bf B}^{\sharp}$.
When $({\sf f}_{\xu,\l})_{\l > 0,\xu \in \Bu} \in {\bf B}^{\sharp}$
relates to $(f_{\l})_{\l >0} \in {\bf F}_q(\fV)$ as in \eqref{equ20}, and if
their primed counterparts are likewise related, then for sufficiently small
$|t|$ and $\xu \in \Bu$, $\xu' \in \Bu'$ one has
$$ \o(\at({\sf f}_{\xu,\l})\a_{t'}({\sf f}'_{\xu',\l})) =
\o_2(T_t^{\stern}R_{\xu}^{\stern}f_{\l} \otimes
T_{t'}^{\stern}R'_{\xu'}\!{}^{\stern}f'_{\l}) $$
for small enough $\l$.
Whence, upon taking 
$$ V = \{(k_0,\ku;k_0',\ku') : (k_0,k_0') \in V_0,\ \ku,\, \ku' \in
\bR^{m-1}\}$$
and\ \,$ h(t,\xu;t',\xu') = h_0(t,t')\underline{h}(\xu,\xu')$, where
$\underline{h}$ is  in $\Coin(\bR^{m-1}\times
\bR^{m-1})$ with $\underline{h}(0) = 1$, and with $h_0$ and
$\underline{h}$ having sufficiently small supports,
 it is now easy to see that
\eqref{equ21} entails the required relation \eqref{equ22}, proving 
\eqref{equ23}, whence \eqref{equ23a} is also established.
\\[6pt]
{\it 3.) } Now we shall show the assumption that $\o^{(+)}_2$ is
smooth at causal separation to imply that also
$\o_2^{(-)}$ and hence, $\o_2$ itself is smooth at causal separation.
The same conclusion can be drawn assuming instead that $\o_2^{(-)}$ 
is smooth at causal separation.
We will present the proof only for the first mentioned case, the
argument for the second being completely analogous.

We define $\cQ$ as the set of pairs of causally separated points
$(q,q') \in M \times M$.
 The restriction of $\o_2$ to $\Coin((\fV \bt \fV)_{\cQ})$
 will be denoted by $\o_{2\cQ}$. By assumption, $\o_{2\cQ}^{(+)}$ has
 empty wavefront set and therefore ${\rm WF}(\o_{2\cQ}) = {\rm
   WF}(\o_{2\cQ}^{(-)})$. Since $(q,q') \in \cQ$ iff $(q',q) \in \cQ$,
 the `flip' map $\r : (q,q') \mapsto (q',q)$
 is a diffeomorphism of $\cQ$. Then 
\begin{equation}
\label{equ21a}
 R: \fV_q \otimes \fV_{q'} \owns {\rm v}_q \otimes {\rm v'}_{q'}
\mapsto {\rm v'}_{q'} \otimes {\rm v}_q \in \fV_{q'} \otimes \fV_q
\end{equation}
is a morphism of $(\fV \bt \fV)_{\cQ}$ covering $\r$.
Thus one finds
$$ [R^{\stern}(f \otimes f')](q,q') = f'(q) \otimes f(q')\,,$$
implying
$$ \o_{2\cQ}^{(-)}(R^{\stern}(f \otimes f')) = \o_{2\cQ}^{(-)}(f'
\otimes f) = - \o_{2\cQ}^{(-)}(f \otimes f')$$
for all $f\otimes f' \in \Coin((\fV \bt \fV)_{\cQ})$. Noting that
multiplication by constants different from zero doesn't change the
wavefront set of a distribution, this entails, with Lemma \ref{lem3}
\begin{equation}
\label{equ24}
{\rm WF}(\o_{2\cQ}^{(-)}) = {\rm WF}(\o_{2\cQ}^{(-)} \lcrc R^{\stern})
= {}^t\!D\r^{-1} {\rm WF}(\o_{2\cQ}^{(-)})\,.
\end{equation}
Now it is easy to check that
$$ {}^t\!D\r^{-1}(q,\xi;q',\xi') = (q',\xi';q,\xi) $$
for all $(q,\xi;q',\xi') \in {\rm T}^*M \times {\rm T}^*M$, and this
implies 
\begin{equation}
\label{equ25}
 {}^t\!D\r^{-1}(\cN_- \times \cN_+) 
 = \cN_+ \times \cN_-\,.
\end{equation}
However, since we already know from \eqref{equ23a}
 that ${\rm WF}(\o_{2\cQ}) \subset \cN_-
\times \cN_+$ and ${\rm WF}(\o^{(+)}_{2\cQ}) = \emptyset$, we see that
${\rm WF}(\o_{2\cQ}^{(-)}) \subset \cN_- \times \cN_+$. Combining this
with \eqref{equ24} and \eqref{equ25} yields
$$ {\rm WF}(\o_{2\cQ}^{(-)}) \subset (\cN_- \times \cN_+) \cap (\cN_+
\times \cN_-) = \emptyset\,.$$
And thus we conclude that $\o_2$ is smooth at causal separation.
\\[10pt]
{\it 4.)} Now we will demonstrate that the wavefront set has the form
\eqref{equ26} for
points $(q,q)$ on the diagonal in $M\times M$, by demonstrating that   
otherwise singularities for causally separated points would occur
according to the propagation of singularities (Prop.\ \ref{pro2}). To
this end, let $(q,\xi;q,\xi')$ be in ${\rm WF}(\omega_2)$ with 
$\xi$ not parallel to $\xi'$. In view of the observation made below
\eqref{equ23a} that we must have $\xi \ne 0$ and $\xi' \ne 0$, we
obtain from Prop.\ \ref{pro2} 
${\rm B}(q,\xi)\times{\rm B}(q,\xi') \subset {\rm WF(\omega_2)}$.
For any Cauchy surface of $M$, one can find 
$(p,\eta;p',\eta')$ in  ${\rm B}(q,\xi)\times{\rm
  B}(q,\xi')$ with $p$ and $p'$ lying on
that Cauchy surface because of the inextendibility of the bi-characteristics. 
Since $\xi$ is not parallel to $\xi'$, one can even choose that Cauchy surface
so that  $p\neq p'$ (if such a choice were not possible, the
bi-characteristics through $q$ with cotangent $\xi$ and $\xi'$ would 
coincide).  But this is in contradiction to the result of
{\it 3.)} since $p$ and $p'$ are causally separated. Hence, only 
$(q,\xi;q,\xi')$ with $\xi=\lambda\xi',\lambda\in\RR$ can be in
${\rm WF}(\omega_2)$.
Applying the constraint
$\xi(\partial^\tau)+\xi'(\partial^\tau)=0$ found in \eqref{equ23a} gives
$\lambda=-1$. Together with the other constraint ${\rm WF}(\o_2)
\subset \cN_- \times \cN_+$ of \eqref{equ23a} we now see 
 that if $(q,\xi;q,\xi')$ is in ${\rm
  WF}(\o_2)$ it must be in $\cR$. 
\\[10pt]
{\it 5.)} It will be shown next that $\o_2$ is smooth  at points $(q,q')$ in
$M\times M$ which are causally related but \textit{not} connected by
any lightlike geodesic: Suppose $(q,\xi;q',\xi')$ were in ${\rm
  WF}(\o_2)$ with $q,q'$ as described. Using  
global hyperbolicity and the inextendibility of the
bi-characteristics, 
we can then find $(p,\eta)$ in ${\rm B}(q,\xi)$ with $p$
lying on the same Cauchy surface as $q'$. As $p$ cannot be equal to
$q'$ by assumption, it must be causally separated from $q'$,
and so we have by Prop.\ \ref{pro2} a contradiction to
{\it 3.)}. Thus, $\o_2$ must indeed be smooth at $(q,q')$.   
\\[10pt]
{\it 6.)} Finally, we consider the case of points $(q,q')$ connected by
at least one lightlike geodesic: Let $(q,\xi;q',\xi')$ be in
${\rm WF}(\o_2)$. To begin with, we assume additionally that $\xi$ is
not co-tangential to any of the lightlike geodesics connecting $q$
and $q'$. As in {\it 4.)} we then find $(p,\eta;p',\eta')$ in 
${\rm B}(q,\xi)\times {\rm B}(q',\xi')$
with $p$ and $p'$ lying on the same Cauchy surface and $p\neq
p'$, thus establishing a contradiction to {\it 3.)}.
  
To cover the remaining case, let $\xi$ be co-tangential to one of the lightlike
geodesics connecting $q$ and $q'$. As a consequence, we find 
$\eta$ with $(q',\eta)\in {\rm B}(q,\xi)$. By {\it 4.)}, we have 
$\eta=-\xi'$, $\xi'\bef 0$, showing $(q,\xi;q',\xi')$ to be in
$\cR$.
\end{proof}
${}\\$
We conclude this article with a few remarks. First we mention that for
the canonically quantized scalar Klein-Gordon field, ${\rm WF}(\o_2)
\subset \cR$ implies ${\rm WF}(\o_2) = \cR$ and thus the two-point
function of every strictly passive state is of
Hadamard form, see \cite{Rad1}. Results allowing similar conclusions
for vector-valued fields subject to CCR or CAR will appear in
\cite{SaV}.

In \cite{Kay.gs}, quasifree ground states have been constructed for
the scalar Klein-Gordon field on stationary, globally hyperbolic
spacetimes where the norm of the Killing vector field is globally
bounded away from zero. Our result shows that they all have two-point
functions of Hadamard form. As mentioned in the introduction,
quasifree ground- and KMS-states have also been constructed for the
scalar Klein-Gordon field on Schwarzschild spacetime \cite{Kay.HH},
and again we conclude that their two-point functions are of Hadamard
form.

In \cite{Fur}, massive vector fields are quantized on globally
hyperbolic, ultrastatic spacetimes using (apparently) a ground state
representation, and our methods apply also in this case.
\\[20pt]
{\Large {\bf Appendix}}
\begin{appendix}
\section{Ground- and KMS-States, Passivity}
\label{app1}
\setcounter{equation}{0}
Let $(\cA,\{\at\}_{t\in\bR})$ be a topological $*$-dynamical system as
described in Section 2. We recall that a continuous linear functional
$\o: \cA \to \bC$ is called a {\it state} if
 $\o(A^*A) \ge 0$ \quad for all $A \in \cA$
and $\o(1_{\cA}) = 1$.
Now let $\hat{f}(t) := \frac{1}{\sqrt{2\p}}\int{\rm
  e}^{-ipt}f(p)\,dp$, $f \in \Coin(\bR)$, denote the 
Fourier-transform. Note that $\hat{f}$ extends to an entire analytic
function of $t \in \bC$. Then a convenient way of defining ground- and
KMS-states is the following:
\\[6pt]
The state $\o$ is a {\it ground state} for $(\cA,\{\at\}_{t\in\bR})$
if $\bR \owns t \mapsto \o(A\at(B))$ is, for each $A,B \in \cA$, a
bounded function and if moreover,
\begin{equation}
\label{equ2}
\int_{-\infty}^{\infty} \hat{f}(t)\o(A\at(B))\,dt = 0\,,\quad A,B \in
\cA\,,
\end{equation}
holds for all $f \in \Coin((-\infty,0))$.
\\[6pt]
The state $\o$ is a {\it KMS state at inverse temperature $\b > 0$}
 for $(\cA,\{\at\}_{t\in\bR})$
if $\bR \owns t \mapsto \o(A,\at(B))$ is, for each $A,B \in \cA$, a
bounded function and if moreover,
\begin{equation}
\label{equ3}
\int_{-\infty}^{\infty} \hat{f}(t)\o(A\at(B))\,dt =
\int_{\infty}^{\infty}
\hat{f}(t+ i\b)\o(\at(B)A)\,dt\,,\quad A,B \in
\cA\,,
\end{equation}
holds for all $f \in \Coin(\bR)$.
\\[6pt]
The state $\o$ is a {\it KMS state at inverse temperature $\b = 0$} if
$\o$ is $\{\at\}_{t\in\bR}$-invariant and a trace, i.e.
\begin{equation}
 \o(AB) = \o(BA)\,, \quad A,B \in \cA\,.
\end{equation}
(Note that we have here additionally imposed
$\{\at\}_{t\in\bR}$-invariance in the definition of KMS state at $\b =
0$. Other references define a KMS state at $\b = 0$ just by requiring
it to be a trace. The invariance doesn't follow from that, cf.\
\cite{BR2}.)
\\[6pt]
 We note that various other, equivalent definitions of ground- and
 KMS-states are known (mostly formulated for the case that $(\cA,
 \{\at\}_{t\in \bR})$ is a $C^*$-dynamical system), see e.g.\
 \cite{BR2} and \cite{Sak} as well as references cited there.

The term `KMS' stands for Kubo, Martin and Schwinger who introduced and
used the first versions of condition \eqref{equ3}. The significance of
KMS-states as thermal equilibrium states, particularly for infinite
systems in quantum statistical mechanics, has been established in
\cite{HHW}.
\\[6pt]
The following properties of any ground- or KMS-state at inverse
temperature $\b >0$, $\o$, are standard in the setting of $C^*$-dynamical
systems, and the proofs known for this case carry over to topological
$*$-dynamical systems:
\begin{itemize}
\item[(i)] $\o$ is $\{\at\}_{t\in\bR}$-invariant
\item[(ii)] $\omega(A\frac{1}{i}\d(A))\ge 0$ \quad for all $A = A^*
  \in D(\d)$\\ (where $\d$ and $D(\d)$ are as introduced at the
  beginning of Section 2).
\end{itemize}
Let us indicate how one proceeds in proving these statements. We first
consider the case where $\o$ is a ground state. Since $\cA$ contains a
unit element, the ground state condition \eqref{equ2} says that for any $A
\in \cA$ the Fourier-transform of the function $t \mapsto \o(\at(A))$
vanishes on $(-\infty,0)$. For $A = A^*$, that Fourier-transform is
symmetric and hence is supported at the origin. As $t \mapsto
\o(\at(A))$ is bounded, its Fourier-transform can thus only be a
multiple of the Dirac-distribution. This entails that $t \mapsto
\o(\at(A))$ is constant. By linearity, this carries over to arbitrary
$A \in \cA$, and thus $\o$ is $\{\at\}_{t\in\bR}$-invariant.

Now we may pass to the GNS-representation $(\varphi,\cD\subset
\cH,\O)$ of $\o$ (cf.\ Sec.\ \ref{sec1} where this object was introduced for
the Borchers-algebra, but the construction can be carried out for
topological $*$-algebras, see \cite{Sch}) and we observe that, if $\o$
is invariant, then $\{\at\}_{t\in\bR}$ is in the GNS-representation
implemented by a strongly continuous unitary group $\{U_t\}_{t\in\bR}$
leaving $\O$ as well as the domain $\cD = \varphi(\cA)\O$
invariant. This unitary group is defined by
$$ U_t \varphi(A)\O := \varphi(\at A)\O\,, \quad A \in \cA,\ t \in
\bR\,.$$
Since it is continuous, it possesses a selfadjoint generator $H$,
i.e.\ $U_t = {\rm e}^{itH}$, and the ground state condition implies
that the spectrum of $H$ is contained in $[0,\infty)$. Therefore, one
has for all $A \in D(\d)$,
 $$ \frac{1}{i}\o(A^*\d(A)) = \langle
 \varphi(A)\O,H\varphi(A)\O\rangle \ge 0$$
and this entails property (ii).

Now let $\o$ be a KMS-state at inverse temperature $\b > 0$. For the
proof of its $\{\at\}_{t\in\bR}$-invariance, see Prop.\ 4.3.2 in
\cite{Sak}. Property (ii) is then a consequence of the so-called
`auto-correlation lower bounds', see \cite[Thm.\ 4.3.16]{Sak} or
\cite[Thm.\ 5.3.15]{BR2}. (Note that the proofs of the cited theorems
generalize to the case where $\o$ is a state on a topological $*$-algebra.)
\section{Proof of Lemma \ref{lem1}}
\label{app2}
We now want to give the proof of Lemma \ref{lem1}. First, we state
some properties of the strict inductive limit of a sequence of locally 
convex spaces, the topology given to $\cB$ being a specific
example. See for example \cite[II, \S4]{Bou} for proofs as well as
for further details.

Let $(E_n)_{n=1}^{\infty}$ be a sequence of 
locally convex linear spaces such that $E_n\subset E_{n+1}$ and the
relative topology of $E_n$ in $E_{n+1}$ coincides with the genuine
topology of $E_n$ for all $n\in\bN$.   
Let $E$ be the inductive limit of the $E_n$, denote by 
$\pi_n:E_n\hookrightarrow E$ the canonical imbeddings of the $E_n$
into $E$ and let $F$ be some locally convex space.
In this situation, we have:  
\begin{itemize}
\item[(a)] $E$ is a locally convex space. 
\item[(b)] A map $f:E\rightarrow
  F$ is continuous iff $f\circ\pi_n$ is continuous for each
  $n$ in $\bN$. 
\item[(c)] A family of maps $(f_\iota)_\iota$, $f_\iota:E\rightarrow
  F$ is equicontinuous iff the family $(f_\iota\circ\pi_n)_\iota$ 
is equicontinuous for each $n$ in $\bN$.
\item[(d)] The relative topology of $E_n$ in $E$ coincides with the
  genuine topology of $E_n$.
\item[(e)] If the $E_n$ are complete, so is $E$.
\end{itemize}
Now we can prove the Lemma:
 
Because of $(e)$, $\cB$ is complete. The 
characterization \eqref{equ1d} of the continuous linear
 forms on $\cB$ is a special
case of $(b)$.

We want to check now that $\cB$ is a \textit{topological} $*$-algebra,
i.e.\ that its $*$-operation is continuous and
 its multiplication $m:\cB\times\cB\rightarrow\cB$ seperately
continuous in both entries:\\       
For $\fv\in\cB$ let $m_{\fv}:\cB\rightarrow\cB$ be the
right multiplication with $\fv$ and denote by $[\fv]$ the smallest
integer such that $\fv_k=0$ for all $k>[\fv]$. By (b), showing
continuity of $m_{\fv}$ amounts to showing the continuity of the maps 
$m_{\fv}\circ\pi_n:\cB_n\longrightarrow\cB_{n+[\fv]}$
where by (d), we can take the topologies involved to be the genuine
topologies of the respective spaces.
As those topologies are direct sum topologies with
\textit{finitely} many summands, the question of continuity can be 
further reduced, finding that $m_{\fv}$ is continuous iff the maps
\begin{equation*}
\Coin(\bt^n\fV)\ni g\longmapsto f\otimes g\hookrightarrow \Coin(\bt^{n+k}\fV)
\end{equation*}
are continuous for all $f\in\Coin(\bt^{k}\fV),\;k\in\bN$. That this is indeed
the case can be checked by taking recourse to the topologies of the 
$\Coin(\bt^l\fV)$.
Therefore, the maps  $m_{\fv}\circ\pi_n$ are
continuous for all $n$, which in turn shows the continuity of right
multiplication on $\cB$.

In the same way, the proof of continuity of the $*$-operation reduces
to showing continuity of the $*$-operation \eqref{equ1} on 
$\Coin(\bt^n\fV)$, which in turn is easy. 
The continuity of the left-multiplication can be proven completely
analogous to that of right-multiplication or inferred from it, using
the continuity of the $*$-operation.  
Therefore, $\cB$ equipped with 
the locally convex direct sum topology is indeed a topological $*$-algebra.

For the proof of the last statements of the lemma, let $\alpha_x$ be a 
$*$-homomorphism of $\cB$ which is the lift of a
morphism $R_x$ of $\fV$ covering $\rho_x$ as stated in Lemma 4.1.
Because of (b) and (d) above, $\alpha_x$  
is continuous iff
its restrictions $\alpha_x\circ \pi_n:\cB_n\rightarrow \cB_n$
are continuous which in turn is the case, iff the maps
$\bt^nR^{\stern}_x:\Coin(\bt^n\fV)\rightarrow\Coin(\bt^n\fV)$
are continuous. That the $\bt^n R^{\stern}_x$ are indeed continuous
follows from density of $\bt^n\Coin(\fV)$ in $\Coin(\bt^n\fV)$
together with the continuity of $R^{\stern}_x$ on $\Coin(\fV)$, the
latter of which can again be checked by inspection of the topology on 
$\Coin(\fV)$.

For the proof of the continuity property \eqref{equ1b}, note that 
$[\alpha_x(\fv)]=[\fv]$ for all $x$, thus it suffices to prove 
the convergence of $\alpha_x(\fv)$ for $x\rightarrow 0$ in the 
topology induced on $\cB_{[\fv]}$. But this
convergence is implied by the assumed smoothness of $R_x$ 
(hence of $R^{\stern}_x$) in $x$ together with (d).  
The proof of \eqref{equ1c} amounts to showing that 
$(\alpha_x)_{\betr{x}\leq r}$ is an equicontinuous set of maps.   
By (c) and (d), the proof can in the by now familiar way be  
reduced to proving equicontinuity of $(R^{\stern}_x)_{\betr{x}\leq r}$  
for some $r>0$.
For the proof of the latter, note that because of the
assumed smoothness of $\rho_x$ in $x$ we find $r>0$ such that 
for each compact set
$K\subset M$ the set $\underset{\betr{x}<r}{\cup}\rho_x(K)$ is 
contained in some other compact
set. Inspection of the topology of $\Coin(\fV)$ shows that this
enables one to find to each given seminorm $\eta$ on $\Coin(\fV)$ 
another seminorm $\eta'$ such that for all $f$
in $\Coin(\fV)$, 
$\eta(R^{\stern}_x f)\leq\eta'(f)$ holds for $\betr{x}\leq r$,
thus proving the desired equicontinuity.      
\end{appendix}

\end{document}